\newif\ifdraft
\newif\ifreview
\newif\ifacm
\newif\ifarxiv
\newif\ifspace
\newif\ifapx \draftfalse
\reviewfalse
\spacetrue
\apxtrue
\arxivtrue
\acmfalse
\ifreview
\PassOptionsToClass{10pt}{article}
\fi
\documentclass[letterpaper,twocolumn]{article}

\ifacm
\else
\usepackage{usenix-2020-09}
\fi
\ifarxiv
\usepackage{newpxtext}
\fi

\usepackage{amsmath}
\usepackage{nicefrac}
\usepackage{xspace}
\usepackage[table]{xcolor}
\usepackage{tabularx}
\usepackage[nameinlink,noabbrev]{cleveref}
\usepackage{graphicx}
\usepackage{booktabs}
\usepackage{comment}
\usepackage{xurl}
\ifacm
\else
\usepackage{titlesec}
\fi

\xspaceaddexceptions{]\}}

\def\Snospace\nobreakspace{\S{}}

\crefname{chapter}{Chapter}{Chapters}
\Crefname{chapter}{Chapter}{Chapters}
\crefname{section}{\Snospace}{\Snospace}
\Crefname{section}{\Snospace}{\Snospace}
\crefname{subsection}{\Snospace}{\Snospace}
\Crefname{subsection}{\Snospace}{\Snospace}
\crefname{subsubsection}{\Snospace}{\Snospace}
\Crefname{subsubsection}{\Snospace}{\Snospace}
\crefname{appendix}{Appendix}{Appendices}
\Crefname{appendix}{Appendix}{Appendices}
\crefname{theorem}{Theorem}{Theorems}
\Crefname{theorem}{Theorem}{Theorems}
\crefname{lemma}{Lemma}{Lemmas}
\Crefname{lemma}{Lemma}{Lemmas}
\crefname{algorithm}{Algorithm}{Algorithms}
\Crefname{algorithm}{Algorithm}{Algorithms}

\ifdraft
\newcommand{\editorial}[2]{{\textcolor{#1}{#2}}}
\else
\newcommand{\editorial}[2]{\ignorespaces}
\fi
\newcommand{\XXX}{\editorial{red}{XXX}}
\newcommand{\xxx}[1]{\editorial{red}{XXX #1}}
\newcommand{\lijl}[1]{\editorial{orange}{lijl: #1}}
\newcommand{\sgd}[1]{\editorial{cyan}{sgd: #1}}

\ifdraft

\else
\excludecomment{outline}
\fi
\ifarxiv
\newcommand{\sys}{Supermassive Blockchain\xspace}
\title{\sys}
\else
\newcommand{\sys}{\textsc{BigBFT}\xspace}
\title{\sys: Scaling Byzantine Fault Tolerance Without Compromised Security}
\fi

\ifreview
\author{Paper \#1976}
\else
\ifacm
\author{Guangda Sun}
\affiliation{
    \institution{National University of Singapore}
    \country{}
    \city{}
}
\author{Jialin Li}
\affiliation{
    \institution{National University of Singapore}
    \country{}
    \city{}
}
\else
\author{
{\rm Guangda Sun}\\
National University of Singapore\\
sung@comp.nus.edu.sg
\and
{\rm Jialin Li}\\
National University of Singapore\\
lijl@comp.nus.edu.sg
}
\fi
\fi

\ifacm
\acmDOI{}
\acmISBN{}
\fi

\ifacm
\else
\titlespacing*{\paragraph}{0pt}{0pt}{1em}
\fi

\newcommand{\eg}{e.g.\@\xspace}
\newcommand{\ie}{i.e.\@\xspace}

\newcommand{\yes}{\textcolor{green}{\ding{51}}}
\newcommand{\no}{\textcolor{red}{\ding{55}}}

\newcommand{\fetch}{\texttt{fetch}\xspace}
\newcommand{\bump}{\texttt{post}\xspace}
\newcommand{\forward}{\texttt{skip}\xspace}

\newcommand{\votec}{\textsc{VoteCheckpoint}\xspace}
\newcommand{\retrieve}{\textsc{Retrieve}\xspace}
\newcommand{\retrieves}{\textsc{RetrieveShard}\xspace}
\newcommand{\pushss}{\textsc{StoreShard}\xspace}
\newcommand{\voter}{\textsc{VoteReconfig}\xspace}
\newcommand{\getsc}{\textsc{RetrieveChunk}\xspace}

\newcommand{\code}[1]{\texttt{#1}\xspace}
 
\begin{document}

\maketitle
\begin{abstract}
Storage scalability is paramount in the era of big data blockchain.
A storage-scalable blockchain can effectively scale out state storage to an arbitrary number of nodes and reduce the storage pressure on each, similar to distributed databases.
Prior research has extensively utilized sharding techniques to attain storage scalability; however, these approaches invariably compromise safety and liveness guarantees.
In this work, we propose a novel state-execution decoupled architecture, and \sys, a novel storage-scalable Byzantine fault tolerance (BFT) protocol that can sustain the deterministic security properties of conventional BFT protocols.
The state management system employs erasure coding to ensure state availability with scalable storage consumption, while the global consensus and execution layers maintain robust security characteristics.
Our evaluation indicates that \sys achieves better storage scalability compared to prior approaches while incurring low network overhead.
\end{abstract} %
\section{Introduction}
\label{sec:intro}

It is common for deployed blockchains to experience growing state sizes over their lifetimes.
The total state sizes of Bitcoin and Ethereum, the two largest permissionless blockchains, have grown by 3.7$\times$ and 7.8$\times$ respectively over the last five years.
These systems require replicating the entire blockchain state on each participant.
This has led to unsustainable storage requirements --- an Ethereum full node now consumes close to 2~TBs of disk space~\cite{eth-requirement} --- and gradual centralization of the network.

Traditional distributed storage and database systems apply \emph{horizontal scaling} to handle the growing state sizes.
A long line of prior research~\cite{ruan2021blockchains,el2019blockchaindb,hong2024gridb,amiri2021sharper,hellings2021byshard,kokoris2018omniledger,dang2019towards,zamani2018rapidchain} has adopted a similar sharding approach to scaling blockchains.
While effective in storage scalability, sharding a blockchain results in fundamentally weaker safety and liveness properties~\cite{shard-model,shard-sybil}.
Sharded blockchains no longer guarantee \emph{deterministic} optimal Byzantine fault tolerance, and their security is particularly vulnerable when committee sizes are small.

Is there a fundamental trade-off between storage scalability and strong blockchain security?
In this work, we argue that this trade-off is due to the tight coupling of transaction execution and state storage in existing architectures.
To execute a transaction (or a sub-transaction), a blockchain node is required to store all the states accessed in the (sub-)transaction.
To provide strong security against any $f$ Byzantine nodes, at least $f+1$ nodes need to execute every transaction, and thus store the necessary state.
When horizontally scaling the system, $f$ also scales, leading to proportionally more nodes to store the state for each transaction.
The total state capacity of the system, therefore, remains constant, despite the larger system sizes.

Inspired by this insight, we propose a new blockchain architecture that decouples transaction execution and state management.
The execution layer takes a totally ordered sequence of transactions from the consensus layer and executes the transactions in order.
The layer, however, is stateless.
A separate state management layer maintains all blockchain state.
It exposes a key-value interface to the execution layer.
When accessing state in a transaction, the execution layer invokes the interface to fetch and update state.
Our decoupled architecture clearly separates the responsibility between the two layers.
Such separation allows us to achieve both storage scalability and strong security.
The execution layer ensures optimal failure resilience by executing each transaction on at least $f+1$ nodes, without compromising scalability due to its stateless nature.
The state management layer only needs to guarantee \emph{state availability}.
We show that state availability can be enforced in a storage scalable manner, using a combination of erasure coding and cryptographic commitments.

We then build \sys, a concrete instance of the state-execution decoupled blockchain architecture.
The \sys design addresses key challenges in the decoupled architecture, including maintaining state safety and availability without compromising storage scalability, and the performance overhead coming from remote state retrieval.
Inspired by log structured merge trees, \sys divides the state management layer into an update table and state checkpoints.
The update table records recent state updates and is fully replicated.
Periodically, \sys runs a checkpointing protocol to take a state snapshot, freeing the update table entries.
To ensure both storage scalability and optimal failure resilience, \sys applies an $RS(3f+1, f+1)$ erasure code on a state checkpoint, and stores encoded chunks on all the nodes.
However, fetching state from a checkpoint requires expensive state reconstruction.
To minimize coding-induced network and computation overhead, \sys also replicates each state shard on a \emph{constant} number of nodes to maintain storage scalability.
All state fetches are served solely by the shard replicas.
When all replicas of a shard fail, \sys re-replicates the shard using reconstructed state from encoded chunks.

We show that \sys is a practical design using large-scale experiments.
We evaluate \sys on up to 100 nodes in both local and wide-area networks.
Our results show that \sys achieves storage scalability similar to sharded blockchains while always matching or exceeding the performance of fully-replicated blockchains.
\sys reduces per-node storage by 10$\times$ at 100 nodes compared to full replication.
Performance wise, \sys achieves up to 2.6$\times$ higher throughput than fully-replicated blockchains due to reduced state maintenance overhead, while incurring negligible latency overhead.
\sys also scales well with increasing node counts, with no significant performance difference from 4 to 100 nodes.
Regarding checkpointing, \sys completes checkpoint creation within 30 minutes for 100 GiB state at 100 nodes.
Last but not least, \sys adds only moderate 2.3\%-30.2\% network traffic for state retrieval, and the overhead is scalable with increasing node counts.

\if{0}
Blockchain state size is undergoing considerable augmentation.
Prevailing permissionless distributed ledgers~\cite{buterin2014next, bitcoin} present state size spanning hundreds to thousands of gigabytes (\cref{fig:bc-state-size}), exposing considerable financial burden to nodes replicating the states and rendering on-chain data storage economically unviable~\cite{balduf2022dude}.
Simultaneously, the implementation of extensive data applications on blockchain platforms is experiencing notable growth~\cite{qi2021bidl,lokhava2019fast,bodkhe2020blockchain,liu2018blockchain,xu2018making,kim2021two,jeong2020design,sun2020blockchain,li2019edurss,shen2019secure,salah2019blockchain,hirtan2018blockchain, sharma2018blockchain}, with state magnitudes potentially attaining terabyte or petabyte scales~\cite{laughlin2014information,rabindrajit2024high,pollard2018eicu,johnson2023mimic,bigtable}, imposing impractical storage requirements at each node if the state is fully replicated.
\lijl{Is there any concrete evidence that these applications on blockchain actually require large state?}
Can state size of a blockchain \emph{scales} with the system size, like traditional distributed databases?
\lijl{Can first bring up traditional databases scale state size horizontally.}

Sharded blockchains, wherein nodes can function by maintaining only a segment of the state, have been the subject of considerable prior research~\cite{ruan2021blockchains,luu2016secure,kokoris2018omniledger,dang2019towards,zamani2018rapidchain,wang2019monoxide,amiri2021sharper}.
Although these solutions scale blockchain state, they introduce compromises between scalability and security, specifically safety and liveness~\cite{pbft}.
Consequently, these approaches fail to uphold the conventional security assurance of Byzantine fault tolerance (BFT)~\cite{bft}, which deterministically tolerates any $f$ Byzantine faulty nodes within a total of $3f+1$ nodes.
\lijl{Can mention that security and liveness concerns have hindered their adoption in practice.}

The trade-off between security and scalability arises from the prevalent \emph{processing sharding} methodology employed by prior research, as depicted in \cref{fig:approach-compare}.
The entirety of processing, encompassing consensus and execution, is partitioned, with only a limited set of nodes (i.e., the committee) involved in transaction processing.
This leads to diminished resilience against safety and liveness attacks.
We observe that only the sharding of the state is crucial for blockchains to achieve storage scalability, while sharding the other components is unnecessary while posing undesirable consequences.
\lijl{Need to improve the previous few sentences. Understanding this concept is crucial for the paper, so we need to clearly explain the different layers and their responsibilities.}
Based on this observation, we introduce a novel approach: we decouple a state management system (SMS) from the execution layer to handle a sharded state, while maintaining global consensus and execution layers.
This innovative methodology achieves a clear \emph{separation of concerns}: the SMS solely addresses storage scalability and availability, which can be effectively managed using advanced erasure coding~\cite{ec} techniques.
Concurrently, the consensus and execution layers can focus on security attributes without concerning about impacting storage scalability.
Meanwhile, the decoupled SMS also introduces the new challenge of supporting execution at different speeds across nodes.
This necessitates managing multiple state versions concurrently and efficiently identifying historical versions that are safe to dispose of.
And, needless to say, SMS must accomplish this while tolerating $f$ faulty nodes.
\lijl{Might need to better explain these challenges. For someone who hasn't read the full paper, these are not easy to understand.}
Luckily, through the careful employment of erasure coding\cite{ec}, we demonstrate that this difficulty can be practically addressed.
\lijl{We do need to summarize the techniques in the introduction.}

\ifspace
\sgd{Updated.}
Based on this framework, we present \sys, a storage-scalable BFT protocol ensuring deterministic safety and liveness properties with up to $\frac{1}{3}$ faulty nodes.
The consensus and execution layers of \sys can integrate existing Byzantine atomic broadcast protocols.
When the execution layer of a node need to access shards to execute transactions, SMS is responsible to bring the shards available to the node, potentially loading them from other nodes.
We present a novel SMS that ensures availability with scalable storage, whose architecture comprises two tiers.
The archive layer utilizes erasure coding to offer a deterministic availability guarantee.
The active layer maintains a fixed number of shard replicas on a randomized node group, ensuring efficient access with high probability.
Both layers achieve storage scalability independently, so as the whole protocol.

We present our preliminary results on \sys.
We analyze the network overhead added by the shard loading, showing it amplifies network traffic only by a constant factor.
We also analyze the redundancy of \sys comparing to the other scaling approaches.
The result shows that, thanks to storage scalability, \sys not only consumes less storage than a fully replicated baseline with any number of nodes exceeding 16, it also improves the redundancy factor by $9.88\times$ compared to the state-of-the-art~\cite{zamani2018rapidchain}. \lijl{need to cite SOTA} \sgd{Done}
In conclusion, \sys achieves substantial enhancements in storage efficiency compared to all prior work with uncompromised security.
\else
Based on this framework, we present \sys, a storage-scalable BFT protocol ensuring deterministic safety and liveness properties with up to $\frac{1}{3}$ faulty nodes.
The consensus and execution layers of \sys can integrate existing Byzantine atomic broadcast protocols, and SMS is responsible to support the retrieval and update of state shards, behaving as the full state is maintained on every node.
We present a novel SMS with two-tier architecture that can guarantee successful retrieval with scalable storage consumption.
We reason about the availability of SMS in detail, and preliminary results show \sys not only consumes less storage than a fully replicated baseline, it also improves the redundancy factor by $9.88\times$ compared to the state-of-the-art~\cite{zamani2018rapidchain}.
\fi

\fi

%
\section{Background and Motivation}
\label{sec:bg}

\subsection{Blockchain Architecture}
\label{sec:bg:blockchain}

In a blockchain system, a set of mutually distrusted nodes collectively implements a replicated state machine~\cite{rsm}.
The system maintains a \emph{world state}, typically an abstract mapping between keys or addresses to arbitrary values.
Clients submit requests in the form of transactions.
Each transaction can read and/or update the world state.
A blockchain provides linearizability~\cite{linearizability} to clients, \ie, the observable behavior of the system is equivalent to a single correct server that executes transactions sequentially while respecting real-time invocation/response ordering constraints.
It offers such guarantee even when nodes may fail or behave arbitrarily, and the network can delay, drop, or reorder messages.

Most deployed blockchains~\cite{bitcoin,ethereum,algorand,sui} follow the \emph{order-execute} model.
In this model, the blockchain nodes collectively run an atomic broadcast~\cite{pbft,hotstuff,honeybadger,narwhal,bullshark} protocol to reach agreement on a total order of client transactions (typically batched in large transaction blocks).
Each node maintains a local copy of the world state, sequentially executes the finalized transactions following the agreement order, and updates its local state based on the transaction logic.
Given deterministic execution and the guarantees of atomic broadcast, world state on correct replicas remain consistent.

Some blockchains~\cite{hyperledger} apply the \emph{execute-order} model, in which nodes execute the transactions first, and then run atomic broadcast to totally order the transaction outcomes.
This model can result in more scalable transaction execution, and permits non-deterministic transaction logic.
Our solution can be applied to both models;
however, we only focus on the order-execute model in this paper given their wider adoption, and briefly discuss the execute-order model in \cref{sec:rel}.

\paragraph{Blockchain storage layer}
All deployed blockchains~\cite{bitcoin,ethereum,sui,algorand} include a storage layer in which each node stores the world state on non-volatile storage medium (\eg, SSD).
The layer facilitates faster failure recovery and increases state capacity of each node.
Instead of reinventing storage solutions, existing blockchains typically reuse well-established persistent key-value stores such as LevelDB~\cite{leveldb} and RocksDB~\cite{rocksdb}.
Accessing and updating the world state then use the \code{get(k)} and \code{put(k, v)} interfaces of the persistent key-value store.

Several techniques have been applied to optimize the blockchain storage layer performance.
Blockchains such as Ethereum store the world state as a Patricia trie~\cite{taocpv3} to reduce storage overhead.
To overcome the slow storage I/O performance, prior systems commonly apply batched state updates and in-memory caching for hot state entries.

\subsection{The Need for Storage Scalability}
\label{sec:bg:motv}

\begin{figure}
    \centering
    \includegraphics[width=0.7\linewidth]{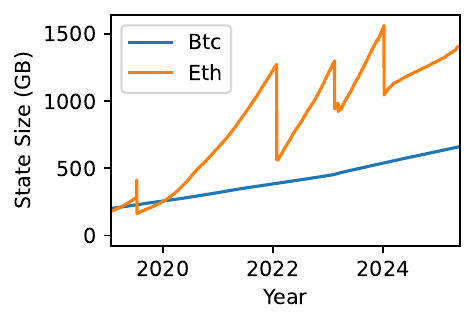}
    \caption{
        Blockchain state size has exhibited rapid growth in recent years.
        The reductions observed in Ethereum's state size correspond to several client software updates incorporating more efficient storage mechanisms and enhanced pruning strategies.
    }
    \label{fig:bc-state-size}
\end{figure}

To maintain feasible storage consumption, existing blockchains offload the ever-growing transaction history to a dedicated archival storage system~\cite{ruan2019fine,feng2024slimarchive}.
However, they still mandate each participating node to store the entire world state in its storage layer.

World state size of deployed blockchains does not remain constant.
New user, wallet, and contract accounts are continuously being created.
It is also common for inactive accounts to remain on a blockchain for years or even decades.
A recent study on Ethereum~\cite{notallstate2025} shows that 63.3\% of the world state has never been accessed since their creation.
Moreover, the state associated with each account can have non-trivial size.
For instance, besides the account balance, each account in Ethereum contains also stores a nonce and multiple hash values;
each contract account can contain 20~KB of code data, tens of KBs of contract storage, and up to a few MBs of calldata.
To understand the storage consumption trend of deployed blockchains, we collected open measurement data of the Bitcoin and Ethereum public networks.
As shown in \cref{fig:bc-state-size}, the world state size (this excludes transaction history) of Bitcoin and Ethereum have grown from 180~GB in 2019 to 660~GB~\footnote{\url{https://www.blockchain.com/explorer/charts/blocks-size}} and 1410~GB~\footnote{\url{https://etherscan.io/chartsync/chaindefault}} respectively in 2025, a 3.7$\times$ and 7.8$\times$ increase over five years.
The blockchain communities have also projected further state explosions in the near future~\cite{eth-state, state-bloat}.

The growing state sizes create several issues for existing blockchain architectures.
Since each participating node needs to store the entire state, the storage capacity requirement increases in tandem with the state size.
As of 2022, the recommended SSD capacity for an Ethereum full node has already reached 2~TBs~\cite{eth-requirement}.
At the same time, SSD price has not seen noticeable decline in the past 18 months~\cite{ssd-price}.
In combination, the cost of operating a blockchain node continues to increase.
This will lead to gradual centralization, where only large organizations with enough storage resources can operate a blockchain node, weakening the resilience of the overall blockchain network.
Earlier participants are forced to leave the network when the state size out-grows their storage capacity.
Moreover, under-utilized storage capacity on any node has zero utility to the rest of the blockchain network.

In permissioned blockchains~\cite{hyperledger}, only authorized nodes can join the network.
However, even in this more managed setting, growing state sizes still present a challenge.
Unlike traditional distributed systems operated by a single organization, nodes in a permissioned blockchain are managed by multiple administrative domains.
Handling larger world state would require \emph{every} organization in the network to upgrade the storage capacity of \emph{all} their nodes.
Any non-compliant organization may result in the halting of the entire system.
This leads to weaker liveness and resilience guarantees, since the blockchain now relies on highly coordinated hardware upgrades from all participating organizations.

Handling larger state is not a unique problem for blockchains.
Prior distributed storage~\cite{gfs, bigtable} and database~\cite{spanner} systems address this challenge by \emph{horizontally scaling} the system to provide higher aggregated storage capacity.
It is, therefore, desirable to enable similar \emph{storage scalability} to blockchain system.
Storage scalability here means that with any additional storage capacity (\eg, more nodes joining or a subset of the nodes expanding their storage), the blockchain can accommodate larger world state.
Besides horizontal scaling, the property also enables nodes with larger capacity to fully utilized their storage resources, while nodes with smaller disks can still participate.
\lijl{Do we need to explain this last point?}

\subsection{Prior Storage Scaling Solutions}
\label{sec:bg:prior}

\paragraph{Sharded blockchains.}
The most prominent approach to blockchain storage scaling is sharding~\cite{ruan2021blockchains,el2019blockchaindb,hong2024gridb,amiri2021sharper,hellings2021byshard,kokoris2018omniledger,dang2019towards,zamani2018rapidchain}.
Similar to distributed databases, these solutions partition the blockchain state into shards, and assign each shard to a subset of blockchain nodes, called a committee.
A committee only stores state for its responsible shards;
it also only processes transactions that access its shards.
For transactions that span multiple shards, sharded blockchains commonly apply a two phase commit protocol.
By adding more committees, the approach can scale the overall storage capacity of a blockchain.

Committee formation is a key design decision for sharded blockchains.
One line of work~\cite{ruan2021blockchains,el2019blockchaindb,hong2024gridb,amiri2021sharper,hellings2021byshard} assigns fault-tolerant clusters as natural committees.
Another line of research~\cite{kokoris2018omniledger,dang2019towards,zamani2018rapidchain} proposes more sophisticated protocols to form committees dynamically, aiming to provide stronger security properties in a permissionless setting.

Sharding improves blockchain scalability, but at the expense of weaker security guarantees.
Traditional BFT-based blockchains have guaranteed safety even when \emph{any} $\frac{1}{3}$ of the nodes are faulty.
However, a sharded blockchain is secure only if \emph{every committee} in the system is secure.
With each committee having only a fraction of the total participants (or total computational power for PoW or total stake for PoS), the likelihood of adversaries taking over a single committee or a committee loses liveness increases.
Mathematical formulations in prior research~\cite{shard-model,shard-sybil} have proven weaker security properties of smaller committee sizes.
Prior solutions all make compromises on the security guarantees or assumptions.
Systems that use randomized committees~\cite{kokoris2018omniledger,dang2019towards,zamani2018rapidchain,algorand} require very large committee sizes for strong probabilistic security, and are vulnerable to biases and predictability in randomness.
OmniLedger~\cite{kokoris2018omniledger} can only tolerate $\frac{1}{4}$ faulty nodes.
RapidChain~\cite{zamani2018rapidchain} assumes a synchronous network model.
The sharded blockchain in \cite{dang2019towards} requires trusted execution environment on all participants.

\paragraph{Stateless blockchains.}
Stateless blockchain~\cite{xu2021slimchain,boneh2019batching,chepurnoy2018edrax,tomescu2020aggregatable,xu2022l2chain,lee2020replicated} is a recent approach that outsources blockchain state and transaction execution to external nodes.
These external nodes submit transaction execution results and succinct proofs to the blockchain nodes, which are only responsible for result validation and reaching agreement on execution order.
The approach allows constant storage overhead on blockchain nodes.
Ethereum plans to adopt similar designs~\cite{eip4844} in its rollup-centric roadmap~\cite{ethereum_scaling_docs,vbuterin2020rolluproadmap}.

Given that the state is no longer maintained by the blockchain nodes, guarantees of a stateless blockchain depend on the security of external nodes.
Liveness issues of the external nodes can impede blockchain progress.
Any state loss in the external nodes halts the entire blockchain.
The blockchain nodes also loses control of transaction ordering to ensure properties such as fairness.
Moreover, the approach does not fundamentally address storage scalability, just pushing the issue to a different layer.
Even for the systems that support state partitioning among external nodes~\cite{xu2021slimchain}, some nodes need to store the entire state to support arbitrary transaction types.

\if{0}

\section{Motivation}

\paragraph{Problem Settings.}
Our work functions as a replicated service for arbitrary state machines~\cite{lamport1978implementation} exhibiting tolerance to Byzantine faults akin to conventional BFT protocols.
The main focus is on the space consumed on participating nodes for storing the \emph{state}, the data component manipulated by the replicated state machine.
\lijl{Could provide more background on what ``state'' means.}
Specifically, \sys ensures both
\begin{description}
    \item[Security]
    \sys ensures safety and liveness properties consistent with conventional BFT protocols~\cite{pbft} with standard settings.
    \item[Storage scalability]
    Given certain amount of the state data, the storage consumption on each node for operating the state is inversely proportional to the number of nodes.
    \lijl{An alternative definition would be: the maximum state size is proportional to the total storage capacity of the system. I think systems people would be more familiar with such a definition.}
    In contrary, state storage in traditional BFT remains the same regardless of the system size.
\end{description}
\sys operates with $3f+1$ nodes and guarantees the aforementioned properties when up to $f$ nodes exhibit Byzantine faults and behave arbitrarily \lijl{Is this assuming all nodes have similar storage capacity?}.
Following numerous prior studies~\cite{pbft,gueta2019sbft,yin2019hotstuff,danezis2022narwhal,spiegelman2022bullshark,luu2016secure,kokoris2018omniledger,dang2019towards,amiri2021sharper,gupta2020resilientdb,wang2019monoxide}, \sys achieves these properties under a partially synchronous network model.
Outside the period of synchrony, the protocol may experience a lack of progress, but safety and storage scalability remain uncompromised.

\sgd{Remember to describe the sharding interfaces in impl.}

\begin{figure}
    \includegraphics[width=0.7\linewidth]{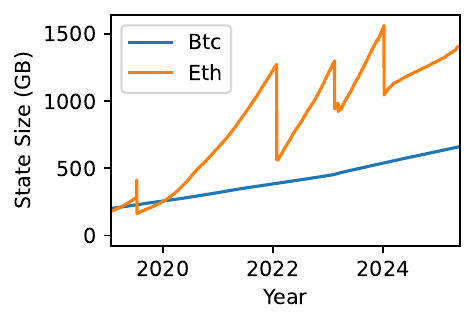}
    \caption{
        Blockchain state size has exhibited rapid growth in recent years.
        The reductions observed in Ethereum's state size correspond to several client software updates incorporating more efficient storage mechanisms and enhanced pruning strategies.
    }
    \label{fig:states-grow}
\end{figure}

\paragraph{Blockchain Storage Consumption.}
The rapidly expanding state sizes of prevalent permissionless blockchains are a well-established phenomenon.
From 2019 to the present, the blockchain states of Bitcoin and Ethereum have grown from approximately 180 GB to 660 GB\footnote{\url{https://www.blockchain.com/explorer/charts/blocks-size}} and 1410 GB\footnote{\url{https://etherscan.io/chartsync/chaindefault}}, respectively (\cref{fig:states-grow}).
While lacking open data on the permissioned settings side, it is reasonable to infer even much larger state sizes based on the significantly higher processing throughput.

\sgd{TODO if possible argue for storing large (instead of many) files on chain.}

\paragraph{The Necessity of Storage Scalability.}
The permissionless blockchains consist of decentralized nodes with heterogeneous storage capacities.
While the state sizes rapidly increase, more and more nodes cannot afford the storage costs and are forced to drop out.
This leads to a gradual centralization of the network, as only nodes with sufficient resources can continue to participate.
At the same time, the nodes with abundant storage resources are not able to utilize their extra storage to benefit the system.
\sys, on the other hand, features storage scalability, is more flexible in configuring the storage load of the nodes.
\sgd{TODO here.}
The increased total number of nodes can effectively alleviate the storage pressure on every node, enabling nodes with limited storage capacity to continue participating in the system.
At the same time, by spawning more virtual nodes, the nodes with abundant storage resources can be further incentivized for contributing more storage to the system \lijl{This last part is not very clear}.

In the permissioned settings, it may seem like the storage overhead is less concerning than in permissionless settings due to the more abundant resources normally found in the deployed environments.
However, because of the static and known membership, these systems cannot replace the participating nodes as easily.
Without storage scalability, the system would require every node to store the entire state.
The system must be halted if any \emph{single} node does not meet the storage requirements.
As more nodes join the network and more data is ingested, the early participants may be overwhelmed.
Thanks to storage scalability, while the state data is growing because of the joining nodes, the storage resources of these joining nodes amortize the storage pressure of these data at the same time.
As the result, the early nodes can enjoy a much slower increasing or even nearly constant storage overhead.
\lijl{When storage capacity is reached, requiring all participating nodes to upgrade their storage. Probably okay when managed by a single organization, but permissioned blockchain usually follows a federated model.}

\lijl{Can also mention that horizontal scaling is a desired property for distributed systems. DAG-based protocols enable horizontal scaling for block dissemination. We are horizontally scaling storage capacity.}

\begin{table*}
\centering
\caption{
    Comparison of blockchain systems.
    \emph{Conventional security} is examined against $\frac{1}{3}$ faulty nodes and partially synchronous network model.
}
\label{tab:compare}
\begin{tabularx}{0.8\linewidth}{XXX}
    \toprule
    Approach                & Conventional security & Scalable storage  \\
    \midrule
    Fully replicated        & \yes                  & \no               \\
    Fault-tolerate clusters & \no                   & \yes              \\
    Randomized committees   & \no                   & \textcolor{orange}{With sufficient nodes}\\
    Concurrent PoW Mining   & \yes                  & \no               \\
    \quad solo mining       & \no                   & \yes              \\
    Stateless blockchain    & \no                   & \textcolor{orange}{Restricted}\\
    \sys                    & \yes                  & \yes              \\
    \bottomrule
\end{tabularx}
\end{table*}

\paragraph{Blockchain Systems with Scalable Storage (\cref{tab:compare}).}
Prior works have proposed various blockchains with low storage overhead~\cite{ruan2021blockchains,el2019blockchaindb,hong2024gridb,kokoris2018omniledger,dang2019towards,zamani2018rapidchain,david2022gearbox,amiri2021sharper,gupta2020resilientdb,wang2019monoxide}.
They can be categorized into two main approaches: \emph{processing sharding} and \emph{stateless blockchain}.
Next, we introduce both approaches in detail and discuss their limitations.

\sgd{Cite~\cite{luu2016secure} in related work as sharded but not storage scalable.}

The processing sharding approach horizontally partitions the state into multiple \emph{shards} and assigns each shard to a subset of nodes, called \emph{committees}\footnote{
    In this line of works \emph{shards} usually refer to the committees, or the \emph{node shards}.
    In this work, we follow the distributed database terminology and use \emph{shard} to refer \emph{data shards}.
}.
\lijl{Since this covers all prior sharding solutions, shall we just call it sharding?}\sgd{This is to distinguish from the data sharding we are about to make.}
Every transaction is processed only by the committee(s) that manages the relevant shard(s).
One important insight is that the committees consist of fixed number of nodes.
As the number of nodes increases, the number of committees also increases, and the state size managed by each committee decreases proportionally, achieving storage scalability.

Based on this approach, \cite{ruan2021blockchains,el2019blockchaindb,hong2024gridb,amiri2021sharper,hellings2021byshard} leverages \emph{fault-tolerant clusters} as the natural committees.
\cite{kokoris2018omniledger,dang2019towards,zamani2018rapidchain} instead design sophisticated \emph{committee formation} protocols to dynamically form committees that are secure with high probability.
\cite{wang2019monoxide} proposes a novel concurrent PoW mining mechanism that enables all committees to share the computational power, securing the committees regardless of node distribution across them.

The stateless blockchains~\cite{xu2021slimchain,boneh2019batching,chepurnoy2018edrax,tomescu2020aggregatable,xu2022l2chain,lee2020replicated} \emph{outsource} the state to external storage nodes.
These storage nodes submit the blocks to the nodes in the system along with the execution results and the proofs that show the correctness of the execution.
Nodes can \emph{attest} the execution results before reaching agreement on them.
As the result, stateless blockchains incur constant storage overhead on the nodes participating the replication.
Ethereum has adopted similar stateless designs~\cite{eip4844} in its rollup-centric roadmap~\cite{ethereum_scaling_docs,vbuterin2020rolluproadmap}.

\paragraph{The Fundamental Trade-off Between Security and Scalability.}
As shown in \cref{tab:compare}, no prior system can ensure conventional security properties with scalable storage.
Fully replicated systems are not scalable in storage consumption, as every node must store a full copy of the state regardless of the number of nodes.
The works taking processing sharding approach generally claims to ensure safety and liveness properties, but each of them introduce various assumptions for ensuring them.
The works relying on pre-determined fault-tolerant clusters~\cite{ruan2021blockchains,el2019blockchaindb,hong2024gridb,amiri2021sharper,hellings2021byshard} can only tolerate $\frac{1}{3}$ faulty nodes within each cluster.
\cite{kokoris2018omniledger} can only tolerate $\frac{1}{4}$ faulty nodes.
\cite{zamani2018rapidchain} assumes synchronous network condition.
\cite{dang2019towards} requires a trusted execution environment.
Also, these works that rely on randomized committees suffer from a lower bound of the committee size (typically 80-600 nodes), meaning that they cannot achieve storage scalability when the number of nodes is less than it.
\cite{wang2019monoxide} claims to achieve the both storage scalability and security, but we remark that the two properties are achieved under different settings.
Following the setting in the original work where the professional mining facilities dominate the mining power, the system can ensure conventional permissionless security but losing storage scalability, as most of the nodes must store the full state to participate into all committees.
On the other hand, the solo miners can enjoy storage scalability by participating only one committee, but they cannot ensure the security properties, as the global $\frac{1}{2}$ faulty power can subvert individual committees in this case.

The stateless blockchains can achieve near-perfect storage scalability, but their unconventional security models introduce untrusted external storage nodes.
The storage nodes are not responsible for ensuring security properties and may behave arbitrarily.
While the internal nodes \lijl{BFT nodes?}\sgd{now?} can ensure safety, the liveness can be simply impeded because of, e.g., the lack of correct storage nodes producing new blocks.
Even worse, the system will permanently halt if the faulty storage nodes lose the state.
Besides, because the nodes give up the right to produce blocks and the control to the order, the systems are also vulnerable to other attack vectors such as fairness.
Additionally, the storage scalability of storage nodes are still not ideal.
\lijl{These weaknesses could be more formalized.}
While \cite{xu2021slimchain} proposes a shard solution for the storage nodes, enabling them to store partial state and process corresponding transactions, there must still be some storage nodes storing the whole state so that all possible transactions can be processed.

The trade-off between security and storage scalability is fundamental in these previous approaches.
In essence, they have to make a decision between two choices: either all nodes process all transactions, or not all nodes process all transactions.
\lijl{What does this mean?}\sgd{done.}
In fully replicated systems, every node processes every transaction, and the system trades off storage scalability for security.
In processing sharding systems, not all nodes process every transaction, enabling storage scalability but trades off security.
Finally, in stateless blockchains, while all nodes participate in the agreement part, none of them execute transactions.
Thus, the system can achieve storage scalability and safety but losing liveness.

\fi
\section{Decoupling State from Execution}
\label{sec:approach}

\subsection{Storage Scalability and Security Trade-off}
\label{sec:approach:tradeoff}

\if{0}
\lijl{Outline:
\begin{itemize}
    \item Tension between storage scalability and strong security
    \item Full state replication guarantees security, but no storage scalability
    \item Sharding scales blockchain storage, but compromises security
    \item This trade-off seems to be fundamental:
    \item To provide strong security, \ie, the system can tolerate any $f$ adversaries, the number of nodes executing each transaction needs to be at least $f+1$.
    \item When adding more nodes to the system, $f$ also increases proportionally
    \item That means, proportionally more nodes need to maintain the state to execute the transactions
    \item The total state capacity of the system remains unchanged
    \item We can similarly show that scaling storage will result in weaker security
\end{itemize}
}
\fi

The discussion in \cref{sec:bg:prior} shows that there is a tension between storage scalability and strong security properties in existing blockchain architectures.
Fully replicating world state on all blockchain nodes guarantees strong security, but the system storage capacity does not scale.
Sharding can scale blockchain storage, but the approach compromises security guarantees.

This trade-off between scalability and security appears to be fundamental.
Conventional BFT-based blockchain security requires the system to tolerate \emph{any} $f$ adversaries within $3f + 1$ total nodes.
To provide such strong guarantee, each transaction needs to execute on at least $f+1$ nodes.
~\footnote{In typical blockchains, each transaction is executed on all nodes, so each node needs to maintain the entire state regardless of the system size.}
When introducing additional nodes to the system, the potential failures to tolerate, $f$, also increases linearly.
The number of nodes that maintains the necessary state to execute the transactions thus also increases proportionally.
Consequently, the total state capacity of the system remains unchanged.
We can similarly show that scaling storage by maintaining constant state size on each node would result in weaker security guarantees.

\subsection{The Case for Decoupling State from Execution}
\label{sec:approach:case}

\if{0}
\lijl{Outline:
\begin{itemize}
    \item This trade-off stems from a deep-rooted architectural decision
    \item In existing blockchain architectures, state and transaction execution are tightly coupled
    \item A node only executes a transaction (or part of a transaction) if it has all the state accessed in the transaction
    \item In traditional blockchain, each node maintains a full copy of the state, and executes all transactions
    \item In sharded blockchains, a node only executes transactions that access its responsible shards
    \item In stateless blockchains, only external nodes execute transactions
    \item Given this tight coupling, it is fundamentally hard to overcome the security and storage scalability trade-off
    \item In this work, we argue for a new blockchain architecture, where transaction execution and state storage are logically decoupled
    \item Clear separation of responsibility: blockchain execution performs transaction statements but is stateless; state storage only maintains blockchain world state, and is agnostic to transaction logic.
    \item This decoupling enables simultaneous storage scalability and the strongest security guarantees.
    \item Execution of each transaction can be done on at least $f+1$ nodes to guarantee deterministic safety and liveness.
    \item State storage is only responsible for ensuring state availability.
    \item Availability is a simpler problem which permits scalable designs.
    \item We highlight two insights that enables scalability.
    \item First, correctness of a state can be independently validated through techniques such as threshold signatures and Merkle trees.
    \item This allows retrieval of state from a single node without safety violations
    \item Second, erasure coding can flexibly reduce the storage redundancy without compromising state availability.
    \item By adjusting the erasure coding parameters, the overall storage capacity can scale with additional storage devices in the system.
    \item Given that the typical working set of a blockchain workload is small, the high computational overhead of symbol decoding can be masked by caching techniques.
\end{itemize}
}
\fi

This trade-off between scalability and security stems from a deep-rooted architectural constraint.
In all existing designs, blockchain state and transaction execution are \emph{tightly coupled}:
A node only executes a transaction (or a partial transaction) if it stores all the state that are accessed in the transaction.
In traditional blockchains, each node maintains a full copy of the state, and therefore executes all transactions;
a node in a sharded blockchain only executes transactions that access its responsible shards;
in stateless blockchains, only external nodes, which maintain state, execute transactions.
Given this tight coupling, it is fundamentally hard to overcome the security and storage scalability trade-off.

In this work, we argue for a radical blockchain design where transaction execution and state storage are logically \emph{decoupled}.
The design results in a clear \emph{separation of responsibility}: Blockchain execution performs transaction statements but is \emph{stateless}; state storage only maintains blockchain world state, but is agnostic to transaction logic.
A surprising benefit of this decoupling approach is that it simultaneously enables storage scalability and the strongest security guarantees.

Execution of each transaction can be done on at least $f+1$ nodes to guarantee deterministic safety and liveness.
\lijl{Can formally define the guarantees of the execution layer}
Since execution no longer maintains state, doing so does not hinder storage scalability.
On the other hand, state storage is only responsible for ensuring \emph{state availability}, \ie, any correct node can eventually retrieve a successfully stored state.
State availability is inherently a simpler problem that permits scalable designs.

We highlight two insights that enables storage scalability while maintaining state availability.
First, correctness of a state can be independently validated using techniques such as hash comparisons or verifying Merkle tree proofs.
This allows retrieving state from a single node without safety violations.
This differs from transaction execution, which requires at least $f+1$ nodes for safety.
Second, erasure coding can flexibly reduce the storage redundancy without compromising state availability.
By adjusting the erasure coding parameters, the overall storage capacity can scale with additional storage devices in the system, while ensuring that any state can be reconstructed in the presence of arbitrary $f$ failures.

\if{0}
From prior research on fully replicated blockchains, sharded blockchains, and stateless blockchains, we observe a common principle: nodes store the state necessary to execute transactions independently.
In other words, the set of transactions a node can execute is determined by the state it maintains locally.
In fully replicated blockchains, every node maintains the complete state and executes all transactions.
In sharded blockchains, nodes only maintain a shard of the state, as they execute transactions that access that shard.
In stateless blockchains, internal nodes do not maintain state because they neither execute nor are able to execute transactions.
Instead, external storage nodes maintain all or part of the state and execute those transactions that can be supported by the state they store.

This ``store what you need for execution'' common theme leads to the dilemma between storage scalability and security guarantees.
On one hand, storage scalability cannot be achieved if the state is replicated to all nodes or a set of nodes whose size is proportional to the total number of nodes.
That means that for each transaction, the state it accesses can only be stored on a limited set of nodes whose number is independent of the total number of nodes.
If only the nodes storing the accessed state can execute the transaction, the number of executing nodes is also limited.
As the number of $f$ faulty nodes increases with the total number of nodes, the faulty nodes can eventually outnumber the executing nodes, making it impossible to achieve security guarantees (without seeking unconventional extensions to the security model).
This fundamental theoretical limitation forces sharded and stateless blockchains to trade off security guarantees in various ways (\cref{sec:bg}) in exchange for storage scalability.
\lijl{This is an important insight. We can make this part even stronger.}

\paragraph{State-execution decoupling.}
On the other hand, we observe that serving consistent state is a fundamentally easier problem than ensuring consistent execution.
The state can be self-verifiable, and receivers can be sure they obtain the correct state even if it is served by single data source.
This is unlike execution which is generally not verifiable without re-execution, so $f+1$ nodes need to execute the same transaction.
Based on this observation, we propose to \emph{decouple state from execution}.

With the proposed approach, each transaction is executed by \emph{all} nodes, regardless of the state they maintain.
The nodes are allowed to maintain only a portion of the state or even no state at all.
We exploit the fact that while the entire state is necessary to execute all possible transactions, each \emph{individual} transaction may not access the entire state, and most transactions in practice access only a small portion of the state.
When executing a transaction, if a node needs to access state that it does not maintain locally, it retrieves the required state from other nodes to complete the execution.
With this state-execution decoupling, we can independently design state management to achieve storage scalability, while all nodes execute all transactions to retain security guarantees.
\lijl{Can relate to other systems designs, e.g., memory is separated from CPU execution logic (CPU only maintains registers and caches), prior disaggregated memory.}
If a transaction accesses a state shard not stored locally, the node retrieves it from other nodes to complete the execution.
By carefully designing the placement of state shards, we ensure that each shard is stored on a limited subset of nodes, thereby achieving storage scalability.
\lijl{A key principle we are arguing here: strong security requires nodes proportional to $f$ to execute the transactions, but availability of data can be done in a scalable fashion.}

\lijl{Another point to make: why do nodes need to maintain some state locally, and how much to store? There are two reasons: state availability and performance. State availability determines the storage lower bound on each node. Storing more state than the lower bound could improve performance.}

\fi
\section{A State-Execution Decoupled Architecture}
\label{sec:arch}

Following our design principle in \cref{sec:approach:case}, we propose a new state-execution decoupled blockchain architecture.
The architecture enables both strong security guarantees and storage scalability.
In this work, we define storage scalability as follows:
Given a certain world state size, the combined storage consumption across all nodes is \emph{independent of} the number of nodes.
The property implies that by adding more nodes to the system, the storage consumption on each node decreases.

\subsection{Architecture Overview}
\label{sec:arch:overview}

\begin{figure}
    \includegraphics[width=\linewidth]{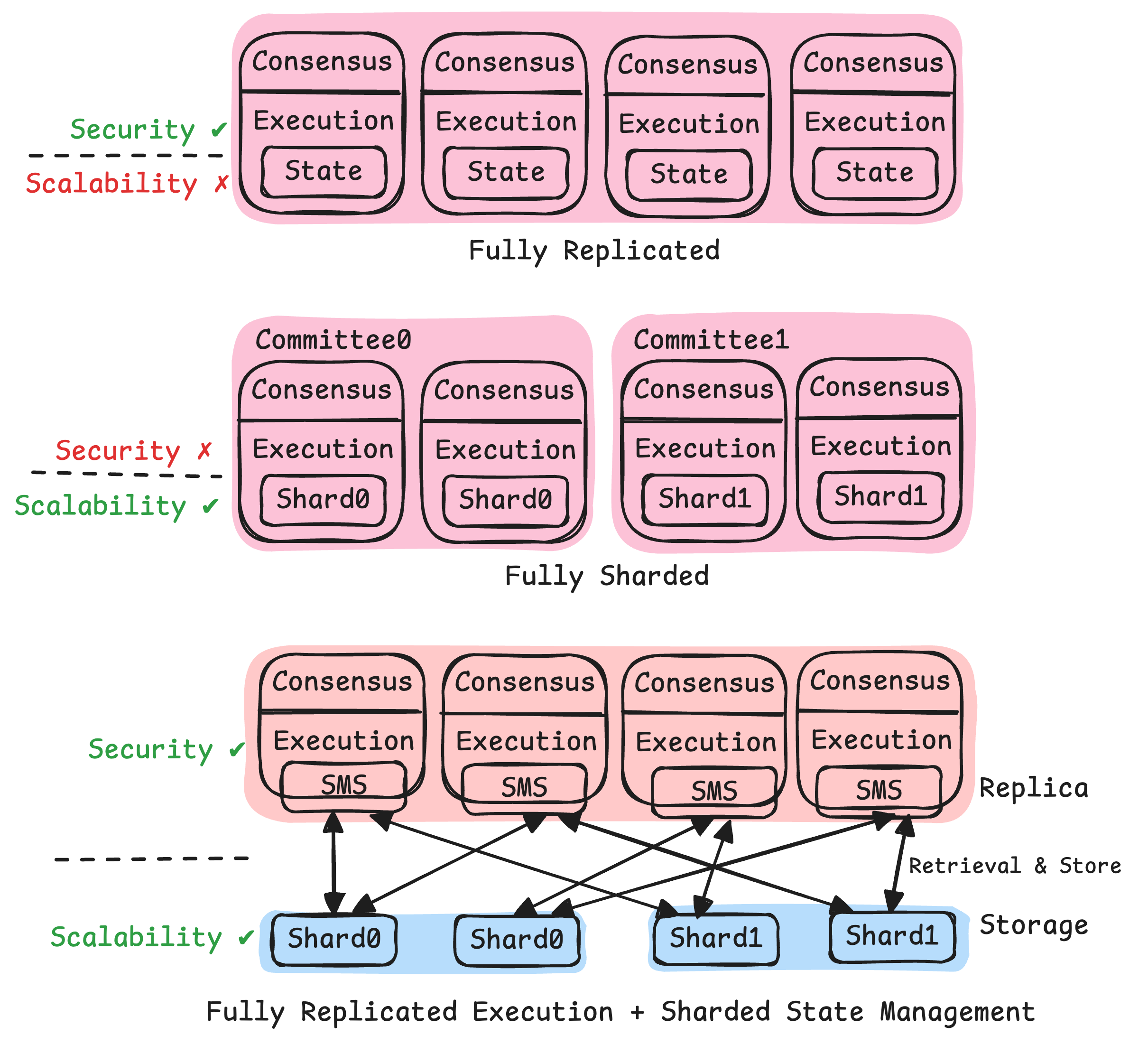}
    \caption{Comparison of blockchain architectures.
    Each rounded rectangle represents a replicated node.
    Each sharded area represents a fault-tolerant unit: a consensus committee can tolerate $f$ faulty nodes with $3f+1$ nodes, while a storage shard can tolerate faulty nodes depending on the redundancy configuration.
    SMS stands for the state management system, the interface of the state management layer.
    }
    \label{fig:approach-compare}
\end{figure}

As shown in \cref{fig:approach-compare}, a state-execution decoupled blockchain is logically partitioned into three layers: a consensus layer, an execution layer, and a state management layer.
The consensus layer and the execution layer follow the traditional design;
the consensus layer establishes a total order of transactions, while the execution layer executes transactions in the agreed order.
However, the execution layer is stateless.
Instead, The state management layer maintains all world state, and exposes an interface to retrieve and update state.
While executing a transaction, the execution layer invokes the interface to read and write world state.

Our state-execution decoupled architecture supports multiple design options.
In this work, we advocate for a design that consists of \emph{logically} separated replica nodes and storage nodes.
A replica node implements protocols in the consensus layer and the execution layer.
It also includes a client-side state management library.
The storage nodes collectively implement the state management layer.
Each storage node only stores a subset of the entire world state.
The two node types are only logical entities, and do not represent physical deployment.
In our basic configuration, each physical node maintains one replica node and one storage node.
This setting offers canonical BFT correctness guarantees: A deployment of $3f+1$ physical nodes can tolerate any $f$ Byzantine failures.
Our model is flexible and supports other configurations.
For instance, each physical node can host more than one logical storage node.
Such deployment allows a subset of participants to contribute more storage space to increase the overall system capacity.
In this setting, the system can also tolerate any $f$ Byzantine physical nodes, but with an additional constraint that the number of storage nodes on the faulty nodes is within $\frac{1}{3}$ of the total storage nodes.
For the remainder of the paper, we only consider the basic setup to simplify discussion.

\subsection{State Management Layer}
\label{sec:arch:layer}

The logically decoupled state management layer is the main research contribution of our work.
The layer exposes a key-value interface, similar to many existing storage systems.
Each transaction generates a set of read and write keys, and invokes the interface to retrieve and update the state.
Such an abstraction is compatible with most blockchain applications, including account-based smart contracts and UTXO-based transactions.

The state managed by the layer is versioned.
Unlike prior versioned data stores, each transaction that mutate keys creates a new version of the entire state.
That is, given an initial state version $v$ and a sequence of read-write transactions, the $i$th transaction reads from state version $v+i$ and updates the state version to $v+i+1$.
Following this model, the state management interface exposes two API calls:

\begin{itemize}
    \item $\fetch_{i}(k)$ --- returns the value of key $k$ in version $i$.

    \item $\bump_{i}([k_1 \mapsto v_1, k_2 \mapsto v_2, \ldots])$ --- creates a state version $i$.
    All keys in version $i$ have the same value as in version $i-1$, except keys $k_1, k_2, \ldots$ which are updated to $v_1, v_2, \ldots$.
\end{itemize}

All replica nodes invoke \fetch and \bump calls to access blockchain state.
Correctness of state management relies on the guarantees provided by the consensus and the execution layers.
Specifically, with a linearizable consensus protocol and correct, deterministic transaction executions, non-faulty replica nodes should generate an \emph{identical sequence} of \fetch and \bump calls.
Given this identical call sequence, the state management layer guarantees the following properties:

\begin{itemize}
    \item \textbf{State Safety}: If a correct replica node issues a $\fetch_{i}(k)$, the call returns the value of key $k$ in version $i$, where state version $i$ reflects the effect of applying $\bump_{1} \dots \bump_{i}$ to the initial state.

    \item \textbf{State Availability}: Every \fetch and \bump call issued by a correct replica node returns eventually. \lijl{Should this only care about \fetch calls?}
\end{itemize}

Since some correct replica node may run arbitrarily slow, naively enforcing state availability would require maintaining potentially unbounded state versions.
To address this issue, the state management layer defines a \emph{state persistence} property.
A state version is \emph{persistent} if the entire state can be retrieved or reconstructed from correct nodes.
We now refine the \fetch and \bump interface.
Suppose the highest persistent state version is $p$, a $\fetch_{i}$ or $\bump_{i}$ call on a correct node with $i < p$ returns $\forward(p)$.
A \forward indicates that the replica node can safely skip all transaction execution until version $p$ without affecting system liveness.

\sgd{If we have space, we should add back the overall proof sketch at here.}

\subsection{Design Challenges}
\label{sec:arch:challenges}

Realizing a state-execution decoupled architecture faces several challenges.
First, state availability requires a state version to survive any $f$ faulty nodes.
The naive approach of storing all state versions on at least $f+1$ nodes, however, violates our storage scalability goal.

Second, in traditional blockchain architectures, a correct node maintains all world state.
The node can therefore trust any locally stored state.
In our decoupled architecture, a replica node may fetch state from remote nodes.
To guarantee state safety, a correct node needs to validate the correctness of remotely retrieved state to tolerate Byzantine behavior.
The node must store some metadata, e.g., the hashes of the state, to validate the retrieved state independently, or it must collect $f+1$ matching states from different nodes, which we already argued violates scalability.
If the node stores only one hash or hashes for fixed number of partitions or encoded chunks of the state, the retrieved state size would be proportional to the entire state size, and transferring such large data will be inefficient.
As such, the storage scheme must not only be scalable, but also enable the nodes to validate the retrieved state with minimal metadata and communication.

Lastly, the decoupled architecture places potential remote state fetches on the critical path of transaction execution.
This introduces additional network traffic and latency penalty compared to a traditional design.
Moreover, the latency overhead also reduces execution throughput, since the transaction execution is serialized.

\if{0}
\paragraph{The size of the retrieved temporary state.}
A natural concern with this design is that, in the worst case, the size of the retrieved temporary state could be as large as the entire state, for instance, if the transactions scan the entire state.
We emphasize that this is a theoretical worst case and is unlikely to occur in practice.
Firstly, as our design will demonstrate into \cref{sec:overview}, nodes retrieve individual key-value pairs from each other instead of transferring the whole state portion the remote nodes maintain.
Thus, the peak size of the retrieved temporary state equals to the \emph{working set}, i.e., the total size of the accessed keys and values, of the executing transactions.
Empirically, most transactions access only an extremely small fraction of the state.
According to the data over 100 days in 2024 from Ethereum, on average each block reads ~1600 slots, about 0.0001\% of the total 151 million slots in the state~\cite{ProperDiskIO_GasPricing_LRUcache_2024}.
In practice, blockchain nodes preserve the working set of one or more recent blocks entirely in memory for performance reasons~\cite{nodereal2022_bsc_cache_hit_rate}.
We follow this practice in our design, assuming trivial memory footprint for the retrieved temporary state without loss of generality.
\lijl{Yes, should provide more concrete evidence and workload citations. Even better if we can do some empirical analysis of available traces.} \sgd{Kind of elaborated.}

\subsection{The Dual Node Architecture for State-execution Decoupling}

Following the state-execution decoupling principle, we propose a dual node architecture where each participant hosts a \emph{replica node} and zero or more \emph{storage nodes}, depicted in \cref{fig:approach-compare}.
The replica node is similar to a traditional BFT node: it contributes to consensus and executes all transactions.
However, it does not maintain the execution state.
\lijl{Shall we name ``execution state'' something else?}
Instead, it reads from and writes to the state through a \emph{state management system} (SMS), which in turn communicates with both the storage nodes of the participant and the ones of other participants.
This architecture practices a clear \emph{separation of concerns}: the storage scalability is fully responsible by the SMS and the storage nodes and will not be impeded by the fully replicated execution.
On the other hand, the replica nodes can ensure the correctness and liveness properties of BFT systems by putting certain requirements on the SMS semantics, which we will elaborate on later in this section.
Because SMS is specialized for state management, the faulty behaviors can be detected by \emph{individual} replica nodes, e.g., by checking the hash of the retrieved data, allowing us to focus on designing a scalable and highly available storage.
This has been the topic of extensive prior research in distributed storage systems.

In this work, we examine the canonical BFT setting, where each participant operates one replica node and one storage node.
This configuration is sufficient to the goal of achieving storage scalability under a conventional BFT security model.
However, the dual node architecture is flexible and can accommodate various configurations.
\sys can tolerate up to $f$ faulty participants that host at most $f'$ storage nodes totally with $3f+1$ participants hosting at least $3f'+1$ storage nodes in total.
This flexibility is especially beneficial in permissionless settings, where participants may have heterogeneous storage capacities.
In such systems, the powerful participants can be incentivized to host multiple storage nodes to enhance the overall storage capacity of the system, while a ``stateless'' participant only needs to operate one replica node to contribute to consensus and execution.
As the result, the powerful participants can make profit from their storage capacities that are potentially even larger than the state size, while the less powerful participants can still participate in consensus and execution without being burdened by storage requirements, maximizing the decentralization of the system.

\subsection{Interfaces and Security Properties of State Management}

The SMS functions as an abstraction layer between the replica node and the storage nodes.
It allows the replica node to read from and write to the state as if it were stored locally.
We adopt a key-value abstraction for the state: the state is represented as a mapping from keys to values, and each transaction accesses a set of keys to read or write their associated values.
This abstraction is compatible with most blockchain applications, including both account-based and UTXO-based models.
\lijl{Can also mention that most distributed databases use such storage interface.}

From the view of state management, we abstract the transaction execution as continuously generating new \emph{versions} of the state and read from those versions.
Initially, there is only the version 0 of the state, containing the genesis state.
Then, more and more versions are created during the transaction execution by modifying the values of certain keys in the current version.
SMS provides interfaces for creating new versions and accessing existing versions of the state.
We define $\bump_{i}([k_1 \mapsto v_1, k_2 \mapsto v_2, \ldots])$ as the interface to create version $i$ of the state as the same as version $i-1$ except that the values of keys $k_j$ are updated to $v_j$ for the key value pairs specified in the call.
Then, we define $\fetch_{i}(k)$ as the interface to read the value associated with key $k$ in version $i$ of the state.
The execution layer can execute transactions with these interfaces.
Starting with $i=0$, it first $\fetch_{i}$ to construct the state needed to execute a transaction or a block of transactions.
Then it executes the transaction(s) deterministically, producing a set of key-value pairs to update.
Finally, it invokes $\bump_{i+1}$ with these key-value pairs, increments $i$ by 1 and continues creating the next version.

Because the execution is replicated, SMS expects all replica nodes to create identical state versions and access the same keys at each version.
By guaranteeing this to SMS, SMS can ensure the following properties.

\paragraph{Correctness.}
On correct nodes, $\fetch_{i}(k)$ correctly returns $k$'s value in version $i$.
This means SMS acts like a local state store to the execution layer.

\paragraph{Availability.}
In the idea case, every $\fetch$ and $\bump$ call returns on all correct nodes.
However, because $f$ nodes may be arbitrarily slower than the others, ensuring this would require maintaining infinite versions of the state.
To address this, we define the \emph{lowest alive version} as the largest version that has been created by at least $2f+1$ correct nodes.
Then, SMS ensures that every $\fetch$ and $\bump$ call at version $i$ returns on all correct nodes if $i$ is no smaller than the lowest alive version at the time of the call.
If $i$ is smaller than the lowest alive version, SMS may return \forward($i'$), hinting the execution layer with the lowest alive version $i'$ and indicating that all versions lower than $i'$ (including $i$) have lost liveness.
The execution layer can then skip executing the transactions that create these lost versions.

\subsection{From SMS Properties to System Security}

Given that the SMS ensures these properties, the BAB protocol ensures its standard properties, and transaction execution is deterministic, we now show that the overall system can guarantee the conventional safety and liveness properties of BFT protocols regardless of the SMS implementation.
The BAB protocol ensures that the execution layer of all non-faulty nodes receives the same sequence of transactions.
Due to deterministic execution, the execution layer of all non-faulty nodes will translate transaction executions into the same sequence of \fetch and \bump calls, satisfying the SMS's expectation.
Then, the availability of SMS ensures that for all transactions, at least $f+1$ non-faulty nodes execute them, i.e., do not receive \forward for the \fetch and \bump calls they invoke during executing these transactions.
Finally, the correctness of the SMS ensures that as long as executions start from consistent pre-states, they yield consistent results and transition to consistent post-states that satisfy the transactions' semantics.
Given that all nodes start from the same initial state, by induction, all non-faulty nodes will produce consistent results for all transactions they execute.
As a result, for each transaction, the client can collect $f+1$ identical responses from executing non-faulty nodes, satisfying the safety and liveness properties of BFT protocols.

\subsection{Challenges of Designing Scalable SMS}

Ensuring availability is challenging.
If a portion of state is stored exclusively on faulty nodes, it may be lost.
However, storing each shard on at least $f+1$ nodes to prevent this results in storage overhead that scales with the number of nodes, thereby undermining storage scalability.

Ensuring correctness is also challenging.
Since the system may contain up to $f$ faulty nodes, these nodes may respond to retrieval requests with incorrect or outdated data.
The node retrieving the state must be able to identify and discard such invalid data to prevent divergence in execution.
A straightforward approach is to collect $f+1$ identical responses from distinct nodes, ensuring at least one correct response.
However, this again leads to storage overhead proportional to the number of nodes.

Last but not least, retrieving state from other nodes introduces additional network traffic and, more importantly, places retrieval on the critical path of execution.
While waiting for retrieval, execution is blocked, and the node cannot process other transactions because execution must remain sequential to ensure consistency.
This leads to increased latency and reduced throughput due to diminished node utilization.
\fi
\section{\sys Design}
\label{sec:design}

\sys is a concrete instance of the state-execution decoupled architecture.
In this section, we give an overview of \sys in \cref{sec:design:overview}, follow by design details in \cref{sec:design:details}.
A full correctness proof can be found in the supplementary material.

\subsection{Overview}
\label{sec:design:overview}

\begin{figure}
    \center
    \includegraphics[width=0.8\linewidth]{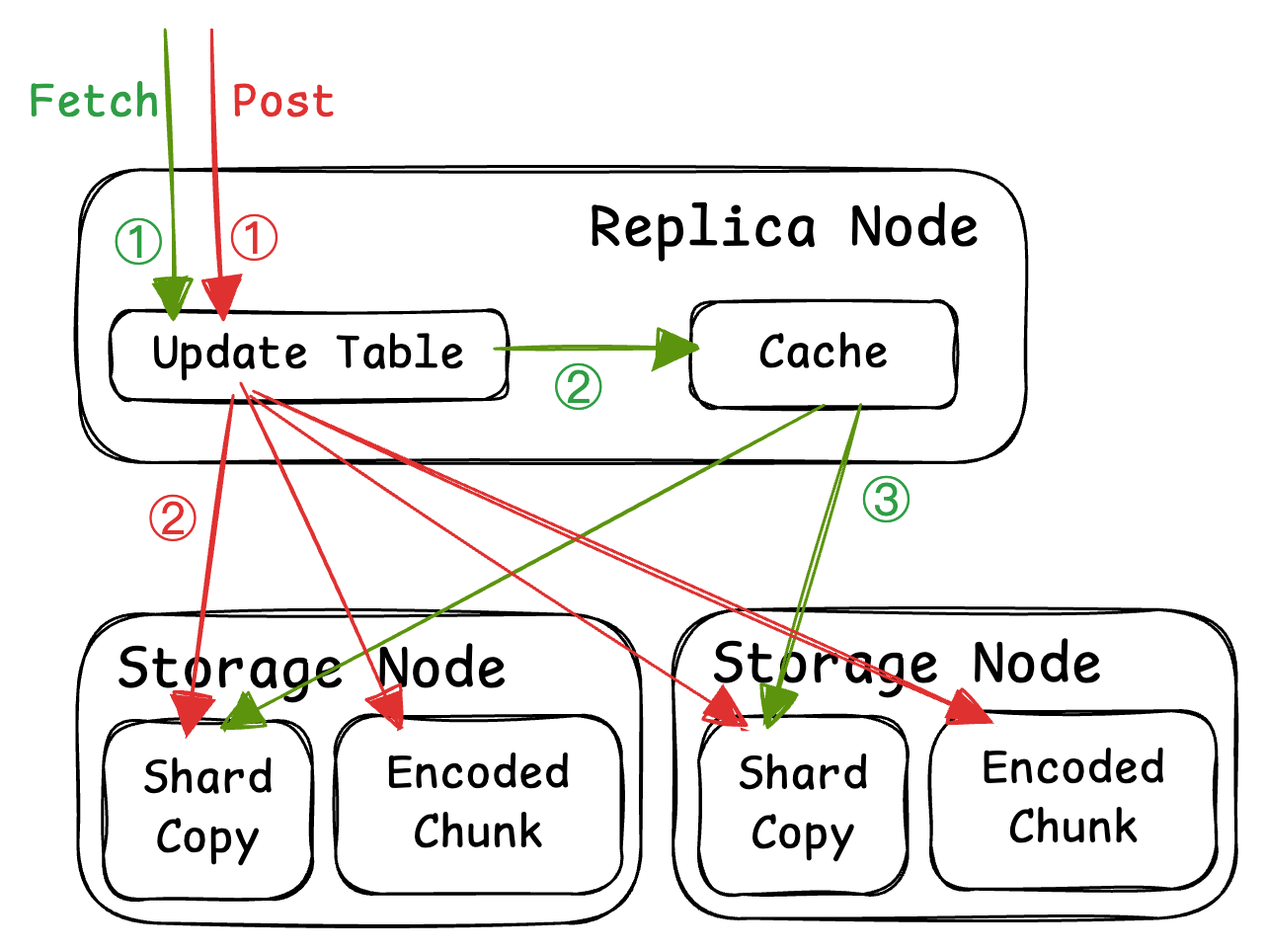}
    \caption{Data flow of \fetch and \bump operations in \sys.}
    \label{fig:flow}
\end{figure}

\sys uses the Bullshark~\cite{bullshark} BFT protocol in its consensus layer.
We pick Bullshark mainly due to its high performance and scalability, but any BFT protocol~\cite{pbft,hotstuff,narwhal,mysticeti} that provides optimal Byzantine resilience and guarantees linearizability can be applied.

\sys divides state management into two layers: an update table and a state checkpoint.
The update table records all recent transaction writes to the state.
This update table is replicated in the state management client-side library on all replica nodes.
Periodically, the storage nodes run a checkpoint protocol to create a snapshot of the current state version.
Once the checkpoint becomes durable, each replica node clears the update table up to the checkpointed version.
\lijl{Explain why replicating the update table does not hurt storage scalability.}
When serving a \fetch, the client-side library first searches the update table for the most recent update of the key.
In case of a table miss, the library queries the latest state checkpoint from the storage nodes.

The \sys storage nodes are responsible for creating, storing, and serving state checkpoints.
To simplify management and to balance workload, \sys partitions state into equal-sized shards.
\lijl{Mention that this ``sharding'' does not cause security issues.}
During state checkpointing, \sys applies an $RS(3f+1, f+1)$ erasure code to encode each shard into $3f+1$ chunks.
These chunks are distributed across all the storage nodes.
Our approach ensures both state availability and storage scalability.
At least $2f+1$ nodes participate in the checkpointing protocol, each storing a distinct encoded chunk for each shard.
Among them, at least $f+1$ shards are stored on correct nodes.
The Reed-Solomon code guarantees that these $f+1$ chunks can recover the state of each shard to guarantee availability.
It is also storage scalable: the storage redundancy level is a constant $\frac{3f+1}{f+1} < 3$, regardless of the system size.

Retrieving state from erasure coded chunks incurs high computation and communication overhead due to the decoding process.
To address this shortcoming, \sys also replicates each shard for fast state retrieval.
To maintain storage scalability, each shard is only replicated on a constant number $r$ of storage nodes, independent of the system size.
These nodes are called the \emph{responsible nodes} of the shard.
The design implies that the replication factor $r$ of each shard can be lower than the maximum tolerable failure $f$.
\sys first applies randomized replication to provide better probabilistic availability guarantees of shard replicas.
When the storage nodes suspect that all $r$ replicas of a shard have lost, they re-elect $r$ responsible nodes for the shard.
These nodes reconstruct a shard copy from the erasure coded chunks, which are guaranteed to be available.
The data flow of \fetch and \bump operations is illustrated in \cref{fig:flow}.

To ensure state safety, \sys organizes each shard into a binary Merkle tree.
The tree is constructed during state checkpointing.
The leaf nodes are the hashes of all key-value pairs in the shard.
The root hash represents a \emph{shard commitment}, and is stored on all replica nodes.
When a replica node retrieves a key, the responsible storage node returns the value and an inclusion proof in the Merkle tree.
The replica node can independently verify the proof using its local shard commitment copy.

After a new state version is checkpointed, the storage nodes can safety garbage collect all previous state checkpoints.
The state management client library also stores the latest checkpoint version $p$.
It returns $\forward(p)$ for any $\fetch_i$ or $\bump_i$ invocation with $i < p$.

\subsection{\sys Design Details}
\label{sec:design:details}

\begin{table}
\centering
\small
\renewcommand{\arraystretch}{1.2}
\begin{tabular}{@{}lp{0.5\linewidth}@{}}
\toprule
\textbf{Message Type} & \textbf{Message Data} \\
\midrule
\multicolumn{2}{@{}l}{\textit{Storage nodes RPC}} \\
\midrule
\retrieve{} request   & round, key \\
\retrieve{} response  & value, inclusion proof \\
\retrieves{} request  & round, shard ID \\
\retrieves{} response & shard data \\
\getsc{} request      & round, shard ID \\
\getsc{} response     & storage node ID, chunk data \\
\pushss{}             & round, shard ID, shard data \\
\addlinespace
\multicolumn{2}{@{}l}{\textit{Atomic broadcast}} \\
\midrule
\votec{}              & round, storage node ID, shard commitments from previous round \\
\voter{}              & epoch, round, replica node ID, nonce \\
\bottomrule
\end{tabular}
\caption{\sys state management protocol messages grouped by channel}
\label{tab:messages}
\end{table}

We now provide design details of \sys.
\Cref{tab:messages} summarizes the protocol messages.
The messages are categorized into two channels according to how they are delivered.
The storage nodes RPC channel is unordered and unreliable point-to-point communication.
\retrieve, \retrieves and \pushss requests are sent by the replica nodes, while \getsc requests are sent by the peer storage nodes.

The atomic broadcast channel is a reliable total order broadcast provided by the underlying BFT atomic broadcast protocol.
Replica nodes initiate these two message types.
A checkpoint signal is formed after the atomic broadcast protocol delivers $2f+1$ matching \votec messages from distinct storage nodes, and a reconfiguration signal is formed after delivering $f+1$ \voter messages with matching epoch and round from distinct replica nodes.
The \votec messages are ordered together with the client requests.
This ensures that all replica nodes checkpoint the same state version.
The \voter messages are ordered so that all nodes reconfigure with the same set of nonce as the seed.
These signals are subscribed by both replica and storage nodes.

Every RPC request and atomic broadcast message contains a \emph{round} number, which indicates the checkpoint round the message belongs to.
The round number starts from 0 at system startup and increments by one after each checkpoint round.
The \voter messages also contain an \emph{epoch} number to indicate the number of reconfigurations that have occurred.

\subsubsection{Randomized Replication Scheme}

The state management system (SMS) first maps each key to a shard ID using a hash function.
Next, for each shard, it determines a \emph{placement}, mapping full shard copies to responsible nodes.
Suppose there are $N$ storage nodes and $k$ shards, we want the placement to 1) replicate each shard to exact $r$ storage nodes, 2) balance the load across all storage nodes, \ie, each storage node is responsible for at most $\lceil rk/N \rceil$ shards, and 3) the probability that each node is responsible for a given shard is uniform across all nodes.

To achieve these goals, we choose a simple Markov-chain algorithm~\cite{kannan1999simple} to generate the placement.
The algorithm starts with an initial placement that assigns each shard $i$ to $r$ (circular) consecutive storage nodes starting with node $i \mod N$.
Next, it repeatedly selects two shards and two storage nodes at random so that each of the nodes is responsible for exactly one of the shards, and swaps the two shards between the two nodes.
After sufficient number of swapping, the placement converges to a uniform distribution.
It also satisfies the replication and load balancing requirements, because the initial placement satisfies them and each swap preserves them.

The randomization procedure is performed with a seeded pseudo-random number generator whose seed is fed from the nonce of the $f+1$ \voter messages.
Each node can independently compute the placement for the shards with the seed during each reconfiguration.
These \voter messages are agreed upon via atomic broadcast, so the placement is consistent across all nodes.
Because at least one of the $f+1$ replica nodes sending the \voter messages is correct, the seed is unpredictable to adversaries.

\subsubsection{Storage Nodes}

The storage nodes are designed as ``dumb'' storage servers.
The stored unit is a shard, and the storage nodes store a shard of a round when it receives $f+1$ matching \pushss messages from distinct replica nodes, ensuring the integrity of the shard.
Then, the stored shard can be retrieved in three different ways: via \retrieve, \retrieves, and \getsc requests.
The responsible nodes respond to all three types of requests, while non-responsible nodes only respond to \getsc requests.

\paragraph{Storage schema.}
We design storage schema to enable efficient responses to \retrieve and \retrieves.
For \retrieve, a node needs to extract the sibling hashes of the key from the Merkle tree to form the inclusion proof.
For \retrieves, the node needs to extract the key value pairs of a specific shard if it is responsible for multiple shards.
To store a shard, the responsible node first sorts the keys in the shard in lexicographical order.
Next, it constructs a Merkle tree whose leaves are the hash of the sorted key-value pairs.
Then it stores each key prepended with the shard ID in big endian, and the value appended with its index in the sorted order.
To respond to a \retrieve, the node first determines the shard ID of the requested key, combines the shard ID with the key to get the value and its index from the storage, then selects the sibling hashes from the Merkle tree with the index.
To respond to a \retrieves, the node seeks to the shard ID and iterates through the keys prefixed with the shard ID.
Each storage node also serializes the (sorted) shard into a byte array and encodes it with the RS$(3f+1, f+1)$ erasure code.
They select one encoded chunk to store according to the storage node ID.
The chunk is stored in a dedicated column family of RocksDB~\cite{rocksdb} with the shard ID as the key.

During reconfiguration, the storage nodes reseed the randomized placement scheme and determine their new responsibilities.
Then, they exchange necessary chunks with each other via \getsc RPCs.
After the new $r$ responsible nodes have reconstructed the shard, they start to respond to \retrieve and \retrieves for the shard.
After checkpointing for round $c$, the storage nodes discard all stored data for rounds before $c-1$ and starts to accept \pushss for round $c$.

\subsubsection{Replica Nodes}

\begin{figure}
    \center
    \includegraphics[width=\linewidth]{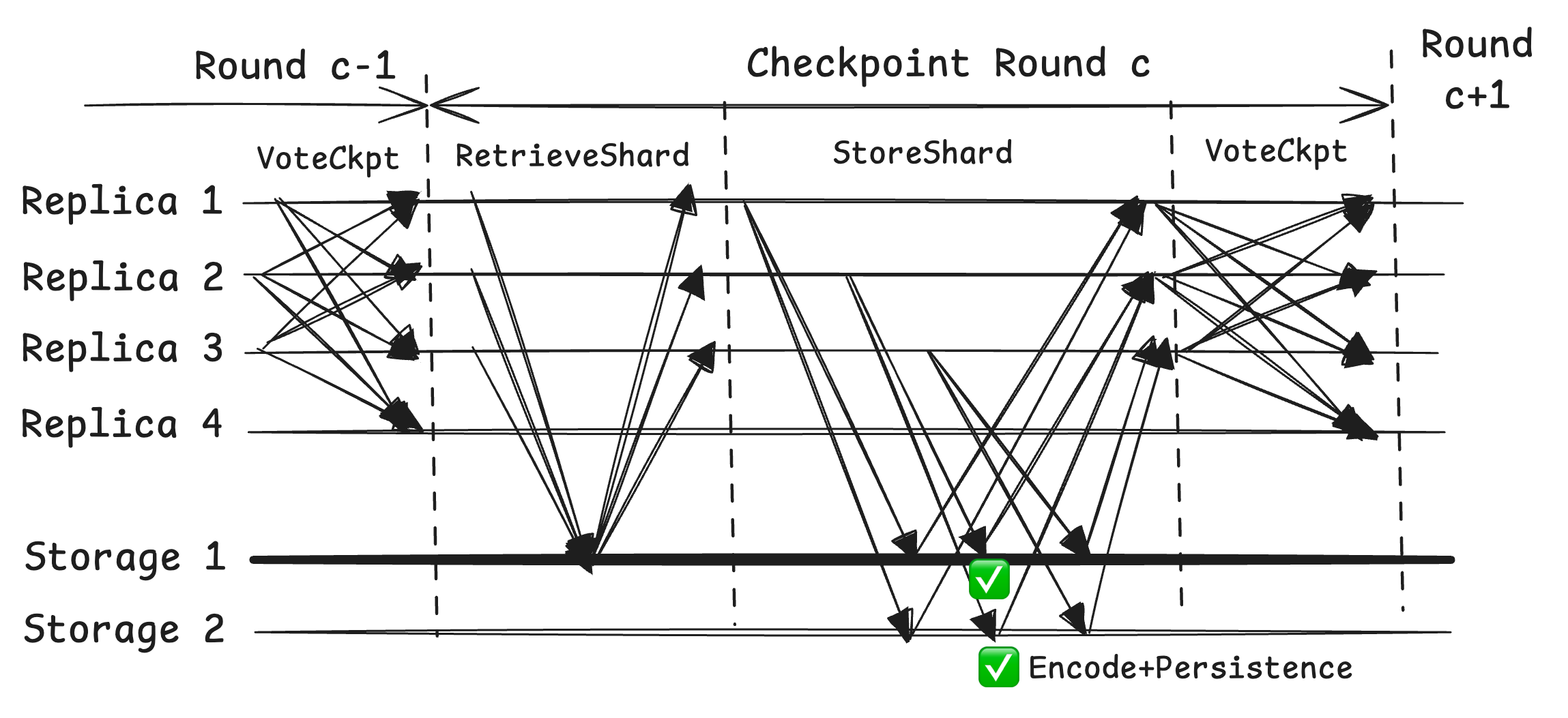}
    \caption{Checkpoint message flow of a shard.
    Storage node 1 is responsible for the shard.
    Replica node 4 is failed.}
    \label{fig:checkpoint-message}
\end{figure}

\begin{figure}
    \center
    \includegraphics[width=0.75\linewidth]{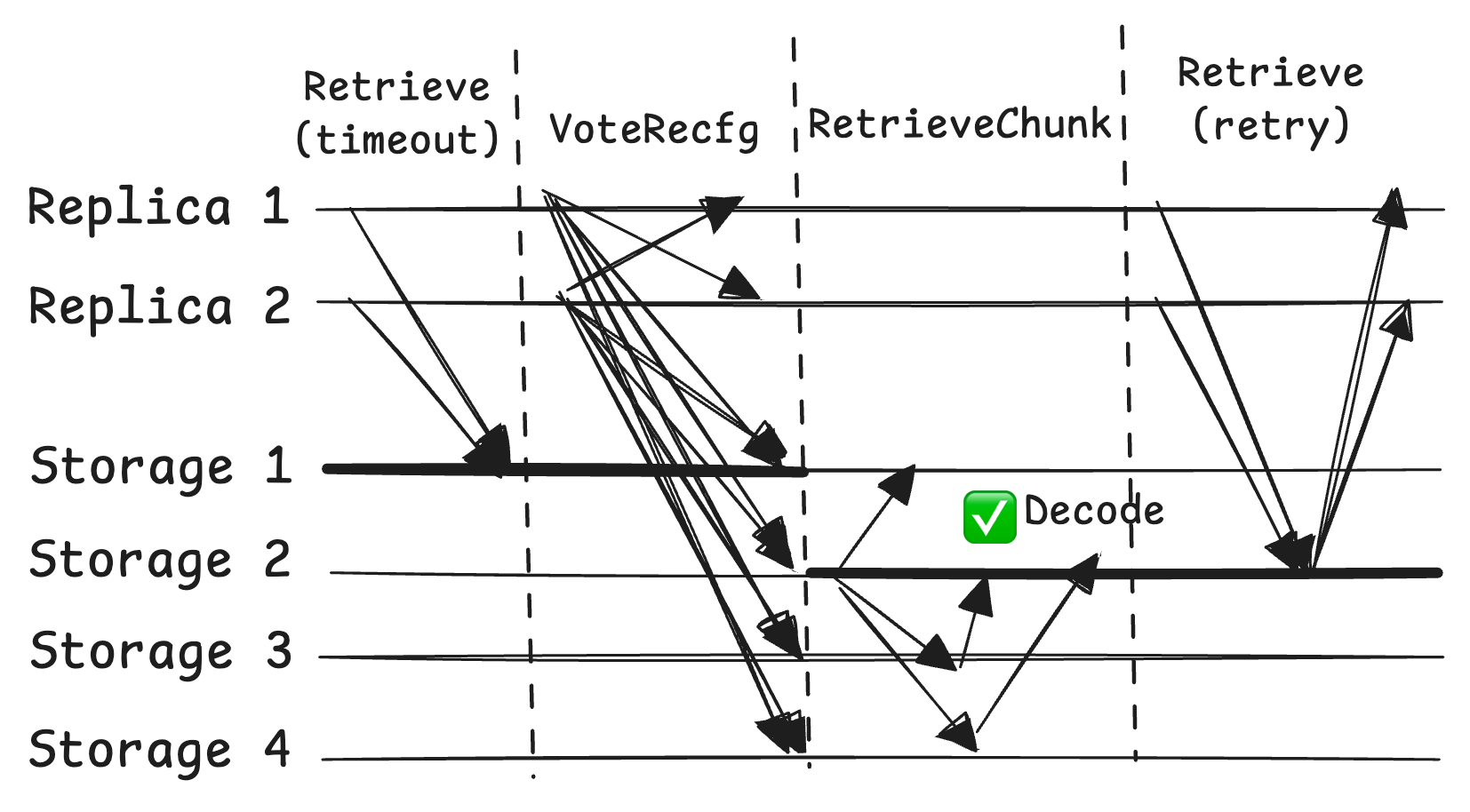}
    \caption{Reconfiguration message flow of a shard.
    Storage node 1 (failed) and 2 are the responsible shards before and after reconfiguration.}
    \label{fig:reconfigure-message}
\end{figure}

The execution layer sequentially processes the ordered messages delivered by the atomic broadcast protocol.
If the message is a client transaction, it executes the transaction and accesses the state by calling \fetch and \bump according to the transaction logic.
Otherwise, it forwards the message to the SMS for processing.
The SMS maintains the current checkpoint round, an update table and a cache (\cref{sec:impl}) on the replica nodes.
On \bump operations, the SMS records the updated keys and their new values in the update table.
On \fetch operations, the SMS first checks the update table for the requested key and returns the value directly if found.
If not found, it checks the cache, and lastly resorts to sending \retrieve requests to the storage nodes on a cache miss.
The \retrieve requests are batched to reduce communication overhead (\cref{sec:impl}).
After getting the responses, the SMS verifies the values and proofs against the shard commitment, updates the cache, and returns the values to the execution layer.

The SMS suspects a liveness failure if the \retrieve requests are not (correctly) responded within a timeout.
In such cases, it increments the epoch and sends \voter messages with its current checkpoint round and a random nonce to the atomic broadcast protocol.

The SMS also records the forwarded \votec and \voter messages and forms the checkpoint and reconfiguration signals.
It sends \votec for the first round of checkpointing at system startup.
On checkpoint signals of round $c$, the SMS performs the checkpointing procedure as follows, except if its current round is less than $c-1$, in which case it skips the checkpointing procedure and \forward to the version corresponds to round $c-1$.
For checkpointing, SMS first snapshots the current update table.
Then, it sends \retrieves requests to the responsible storage nodes, retrieves the shards, and verifies them against the shard commitments in the \votec messages.
After that, it applies the relevant updates in the update table snapshot to the retrieved shards and computes the new shard commitments.
Next, SMS sends \pushss messages with the new shards to all the storage nodes and waits for $2f+1$ acknowledgments.
After the SMS stores all the shards for round $c$, it discards the snapshot of the update table, and sends \votec messages for round $c+1$ with the new shard commitments.
\Cref{fig:checkpoint-message} shows the protocol diagram of state checkpointing.

On reconfiguration signals, the SMS reseeds the randomized placement scheme, determines the new responsible nodes for each shard, and sends the following \retrieve and \retrieves requests accordingly.
The protocol diagram of reconfiguration is illustrated in \cref{fig:reconfigure-message}.
\section{Implementation}
\label{sec:impl}

We built a \sys prototype to validate its design and performance.
Replica nodes run a partial-synchronous Bullshark~\cite{spiegelman2022bullshark} implementation from scratch and use RocksDB for storage.

\paragraph{Shard sizing and counts.}
The number of shards $k$ must balance memory fit (state size/$k$), load balancing ($k \gg N$ storage nodes), and management/checkpoint overhead (more shard hashes in \votec and longer checkpointing).
We therefore use $k=1000$, which fits our evaluation, and reconfiguration can adjust $k$ and $N$ as needed.
We set $r=7$, yielding 99.9\% data availability.

\paragraph{Replica-local caching.}
We disable RocksDB’s cache on storage nodes and instead use an in-memory LRU cache at replicas.
This keeps the replica memory footprint comparable to full replication while avoiding the combined network and verification cost when storage nodes would otherwise serve cached data.
We size this cache to match the RocksDB cache size in the baseline systems.

\paragraph{Batched fetching and retrieval.}
Sequential fetch, execute, and post collapses throughput under high latency. Each transaction waits one retrieval RTT, which could be hundreds of milliseconds in WAN settings, leading to single-digit TPS and idle CPU and network resources.
So \sys batches: the execution layer processes blocks of transactions, concurrently \fetch all needed keys, SMS batches \retrieve RPC requests, and the execution layer issues one \bump for all block updates.
Thousands of transactions thus become ready after a single round trip, greatly improving throughput.
Also, when the same key is accessed multiple times in a batch, SMS deduplicates retrievals to avoid redundant network and disk work.
Batched fetching requires knowing keys ahead of execution (via static analysis or annotations, e.g., Ethereum access lists~\cite{heimbach2024dissecting}).
The YCSB and UTXO applications used in our evaluation (\cref{sec:eval}) have keys known a priori.
Read-after-write within a batch is preserved by maintaining an intermediate intra-batch partial state.
Pipelining fetch and execution could further improve utilization for heavier workloads (e.g., EVM), but we omit it because fetching already dominates in our lightweight evaluations.

\paragraph{Primary copies.}
Each key resides on $r$ nodes, and naïvely retrieving from all $r$ multiplies network and disk cost.
We therefore designate one responsible node as primary: when its trusted replica issues a \retrieve RPC request for a key, the primary pushes the value to other replicas.
Replica nodes first hold the \retrieve request briefly to await this push and only then fall back to fetching from all $r$ if the push does not arrive within the timeout.
To avoid faulty primaries degrading performance, primaries also push to other responsible nodes, which monitor and deterministically replace misbehaving primaries.
Replacements keep pushing for a short grace period even after the faulty primary recovers, ensuring continuity and bounding performance impact.

\paragraph{Offload shard construction to storage nodes.}
The baseline design has replicas construct new shards during checkpointing while storage nodes expose simple store and retrieve operations.
We can offload construction by having replicas send update tables to their trusted storage nodes (replicas without trusted storage nodes skip checkpointing).
Leveraging trusted direct transmission, storage nodes do not need to collect matching update tables from multiple replicas.
Storage nodes then drive checkpointing independently of replicas: they load or send \retrieves for prior shards, apply updates, construct new shards, compute hashes, and broadcast \votec messages to finalize the checkpoint.

\begin{figure}
    \includegraphics[width=\linewidth]{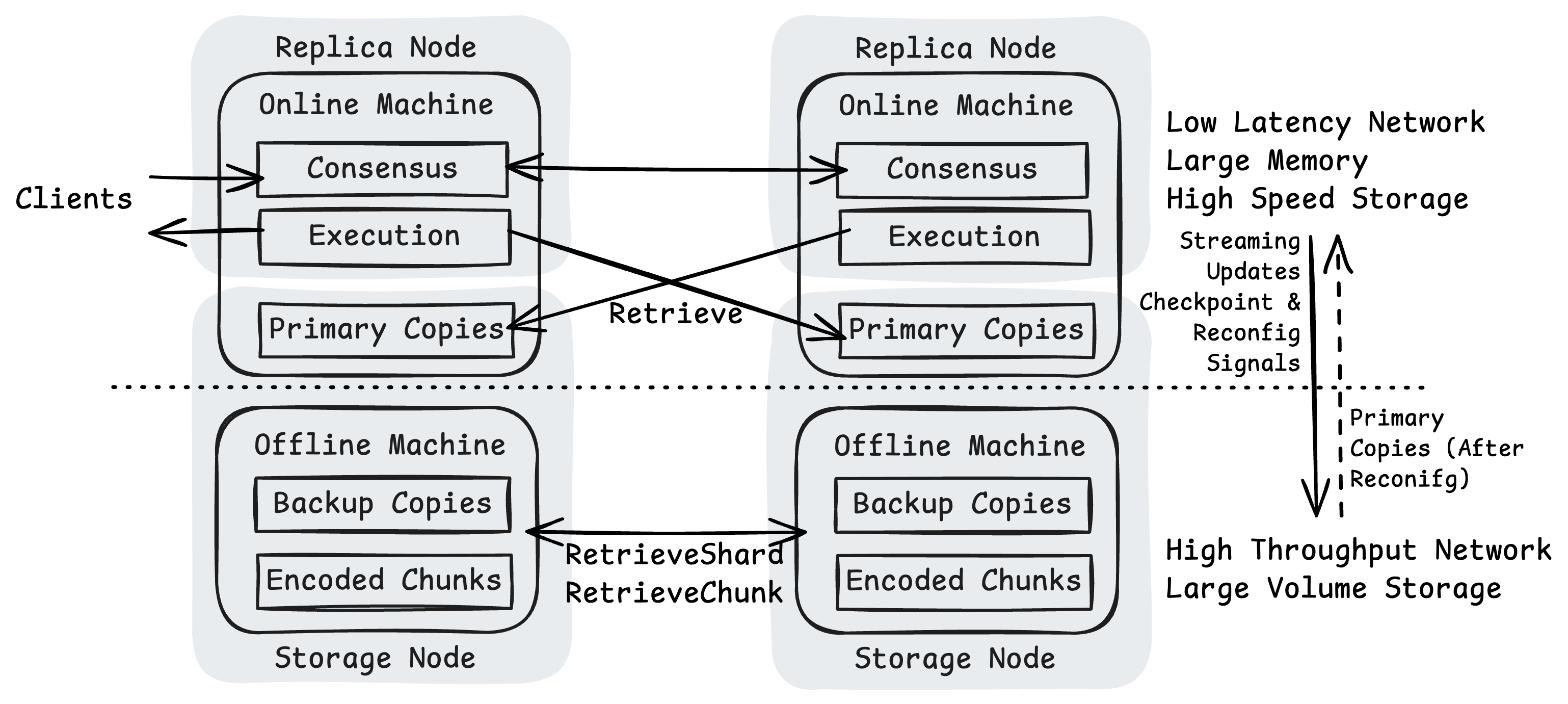}
    \caption{Separation of online and offline machines to isolate performance between retrieval and checkpointing.
    The rounded rectangles represent physical machines, while the shaded area represents logical replica nodes and storage nodes.
    }
    \label{fig:online-offline}
\end{figure}

\paragraph{Performance isolation between retrieval and checkpointing.}
Replica and storage nodes are logically separate but may be co-located on the same physical machines for resource efficiency.
However, co-location can cause performance interference: \retrieve RPCs are latency-sensitive with small payloads, whereas checkpoint \retrieves are bulk transfers that require high throughput but are insensitive to latency.
Deployments optimized for one may degrade the other.

We therefore propose splitting each participant into \emph{online} and \emph{offline} machines connected by a trusted channel (\cref{fig:online-offline}).
The online machine hosts the replica node and primary copies, handling latency-sensitive \retrieve requests and consensus communication.
As shown in \cref{sec:eval}, \retrieve traffic remains small relative to consensus, so the online machine's network demands resemble those of a fully-replicated BFT node.
It requires moderate storage (only primary copies, potentially fitting in memory with aggressive caching) and low-latency network.
The offline machine hosts remaining storage and handles bulk checkpoint transfers, prioritizing high-bandwidth network and large storage capacity.
A typical deployment pairs a high-performance server (online) with a cost-effective storage server (offline).
\section{Evaluation}
\label{sec:eval}

\subsection{Setup}

We evaluate \sys on AWS using up to 70 \texttt{m5ad.2xlarge} nodes (8 vCPUs, 32 GiB memory, 300 GB NVMe SSD, 10 Gbps bandwidth) deployed within a single availability zone across 4--100 replicas with faulty nodes simulated by silence.
RocksDB is deployed on XFS-formatted SSD storage with discard option enabled, following RocksDB's benchmark practice.
For wide-area experiments, we apply artificial latency from a public dataset~\cite{Reinheimer2020} of ping times between 247 cities worldwide (average 184ms, max 400ms), with each node randomly placed in a different city.

We deploy 20 \texttt{c5a.2xlarge} client instances and run different numbers of concurrent close-loop client requests to generate workload.
We run experiments for 30 seconds after warm-up (10s LAN, 20s WAN), reporting average throughput and median latency.

\sys is compared against a fully-replicated BFT baseline and a sharded BFT system, both using Bullshark for consensus.
For fully-replicated and \sys, we vary $f$ from 1--33 (4--100 replicas); for sharded BFT, we fix $f=3$ per shard and vary shard count 2--10 (100 replicas total).
The sharded system orders requests via unified atomic broadcast, then each cluster executes its portion, ensuring aligned consensus overhead.
Sharded BFT has weaker execution-layer fault tolerance: clients collect only $f+1$ matching responses per cluster.
We implement the batched concurrent \fetch optimization for all systems.

We evaluate two applications: (1) a key-value store with 100M pairs (16-byte keys, 1000-byte values, 100 GiB total) following YCSB workload B~\cite{cooper2010benchmarking} (95\% reads, 5\% writes), and (2) a UTXO-based cryptocurrency with 100M UTXOs (32-byte IDs, 4-byte indices, similar to Bitcoin's current size~\cite{MempoolUTXOReport2025}).
For UTXO, clients repeatedly transfer a single initial UTXO to random addresses.
In the sharded BFT system, cross-shard transfers (where input and output UTXOs are managed by different clusters) require a client-coordinated two-phase commit~\cite{kokoris2018omniledger}: clients collect signed locks from input clusters, then submit with locks to both input and output clusters.
YCSB is read-heavy (1000 bytes per operation), while UTXO is write-heavy (one read, two writes per transfer).

\paragraph{Skewness and caching.}
Blockchain state access is highly skewed: 20\% of Ethereum accounts drive 92\% of transactions~\cite{krol2021shard}, the top 0.1\% of state accounts for 62\% of accesses~\cite{lin2025parallelevm}, and 63.3\% is never accessed~\cite{notallstate2025}.
Consequently, caching is highly effective; BNB Smart Chain achieves 90\% hit rates with basic tuning and 99\% with prefetching~\cite{nodereal2022_bsc_cache_hit_rate}.

We model skewed access using Zipfian distribution ($\alpha=1.24$) where the top 1\% keys account for 98\% of accesses.
We configure 4 GiB local cache preloaded with 1M most-accessed keys, achieving 90\% cache hit rate.
We also evaluate performance across skewness factors $\alpha \in [0.0, 1.24]$.

\subsection{Storage Overhead}
\begin{figure}
    \center
    \includegraphics{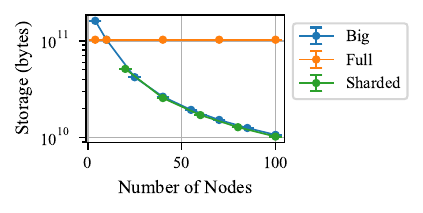}
    \caption{Storage overhead per node across cluster sizes.}
    \label{fig:nodes-storage}
\end{figure}

We prefill various numbers of nodes with the 100M key-value pairs YCSB state and measure the storage overhead using the \texttt{du} command.
Results in \cref{fig:nodes-storage} show that fully-replicated BFT stores 100 GiB per node regardless of cluster size, while \sys achieves storage scalability matching sharded BFT.
At 100 nodes, \sys uses 9.82 GiB per node: 7 GiB from $r=7$ full shard copies and 3 GiB from $3\times$ redundant erasure-coded chunks, confirming that storage overhead decreases with cluster size.

\subsection{Performance Under Different Network Conditions}

\begin{figure}
    \center
    \includegraphics{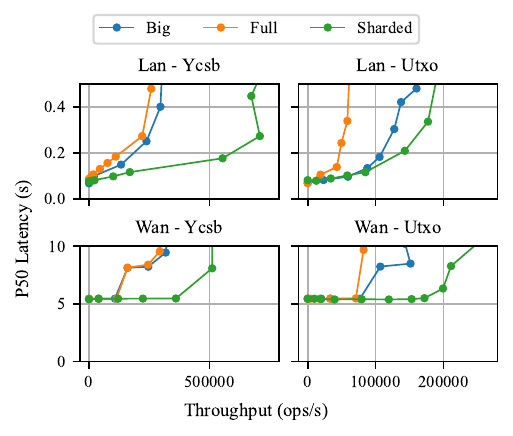}
    \caption{Throughput and latency for YCSB and UTXO under LAN and WAN.}
    \label{fig:tput-latency}
\end{figure}

We benchmark all three systems with 100 nodes under both LAN and WAN conditions, showing throughput and latency results in \cref{fig:tput-latency}.
\sys maintains consistent latency across LAN and WAN (\textasciitilde670ms and \textasciitilde5.44s respectively, matching fully-replicated BFT) because SMS adds only a single-trip overhead via primary pushing (\cref{sec:impl}), which is negligible compared to the consensus latency dominated by Bullshark.

On throughput, \sys outperforms fully-replicated BFT in both applications.
For YCSB, cache hits dominate in both systems; misses incur network retrieval but save disk I/O, which is more expensive than remote access and lightweight verification.
For write-heavy UTXO, \sys's trivial update-table mutations outweigh full-state maintenance costs, achieving $2.6\times$ throughput.

Sharded BFT achieves highest throughput: $2.24\times$ for YCSB (limited by workload skewness causing load imbalance) and $1.5\times$ for UTXO (limited by cross-shard communication and two-phase commit).

\subsection{Performance Under Different Skewness Factors}

\begin{figure}
    \center
    \includegraphics{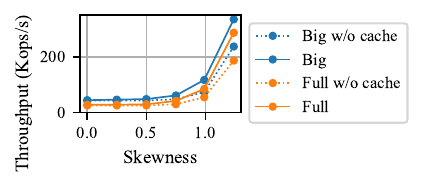}
    \caption{YCSB throughput across Zipfian skewness factors and cache sizes.}
    \label{fig:skew-tput}
\end{figure}

We compare the saturated YCSB throughput of \sys and fully-replicated BFT under different Zipfian skewness factors ranging from 0.0 (uniform) to 1.24 (highly skewed).
Results in \cref{fig:skew-tput} show that \sys experiences significant performance degradation at lower skewness: at uniform distribution, \sys achieves only 13.7\% of highly-skewed throughput, though still exceeding fully-replicated BFT (12.1\%), confirming comparable per-read overhead even when most \fetch operations require remote access.
Disabling cache reduces throughput by 29.3\% (\sys) and 34.9\% (fully-replicated) at $\alpha=1.24$; batch deduplication in \fetch (\cref{sec:impl}) minimizes cache-free performance loss.

\begin{figure}
    \center
    \includegraphics{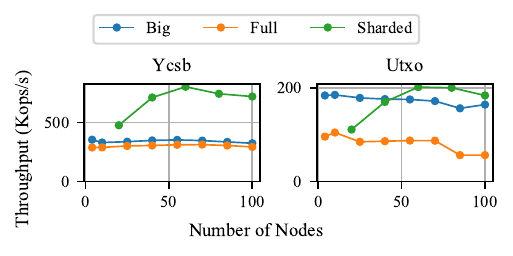}
    \caption{Saturated throughput across node counts.}
    \label{fig:nodes-tput}
\end{figure}

We vary the number of nodes from 4 to 100 and measure the saturated throughput of all three systems for both applications.
As shown in \cref{fig:nodes-tput}, \sys and fully-replicated BFT scale consistently: while storage nodes must serve retrieval requests to more nodes as cluster size grows, more storage nodes share this load, keeping per-node overhead nearly constant.
Sharded BFT shows sublinear scaling, saturating at 60 nodes (6 shards) because consensus overhead persists, skewed YCSB creates load imbalance, and UTXO's cross-shard coordination dominates.

\subsection{Checkpoint Latency}

\begin{figure}
    \center
    \includegraphics{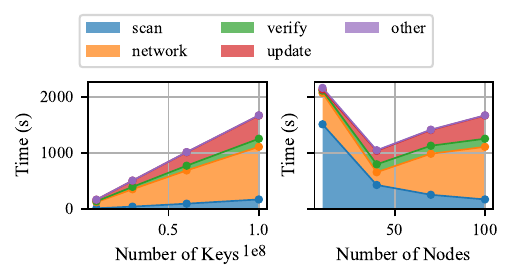}
    \caption{Checkpoint creation time: (left) varying state size at 100 nodes; (right) varying node count at 100 GiB.}
    \label{fig:checkpoint}
\end{figure}

We measure the time taken to create a new checkpoint with zero-sized update table by running offline machines against the prefilled state.
We vary both state size and node count, showing the breakdown in \cref{fig:checkpoint}.
With 100 nodes and 100 GiB state, checkpoint intervals reach 28 minutes, acceptable for practical blockchains.
All stages scale linearly with state size: total time increases from 168s for 10 GiB to 1670s for 100 GiB, with scanning (10.5\%), retrieval (51.0\%), verification (8.8\%), and construction/encoding (29.6\%) maintaining consistent proportions.

From 4 to 40 nodes, checkpoint time halves due to parallel scanning and retrieval.
Beyond 40 nodes, increased per-node remote retrieval (as storage per node shrinks) dominates, lengthening total time.
This suggests that a minimum of 40 nodes provides a good balance for checkpoint latency in \sys.

\subsection{Network Scalability}

\begin{figure}
    \center
    \includegraphics{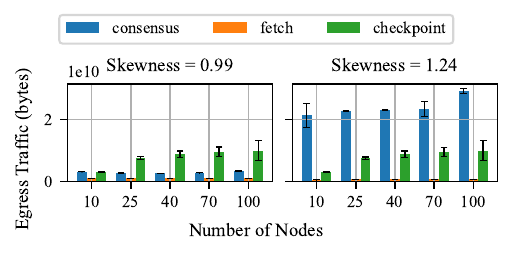}
    \caption{Per-node network traffic for consensus, retrieval, and checkpoint across cluster sizes.}
    \label{fig:nodes-network}
\end{figure}

We measure the proportions of network traffic contributed by each component of \sys and how they scale with node count.
We perform measurements with 4 to 100 nodes under YCSB workload at skewness factors 1.24 and 0.99, adopting a realistic 500 MiB update table size budget (meaning 20M and 2.2M operations per checkpoint period respectively).
We measure egress network traffic of each node during 1/10 of the checkpoint period.

Results in \cref{fig:nodes-network} show that consensus and retrieval traffic remain constant per node as cluster size grows; checkpoint traffic increases as nodes store less state and retrieve more remote shards, but is bounded by the total state size.
Retrieval adds modest overhead compared to fully-replicated BFT, which only incurs consensus traffic: 2.3\% at high skewness and 30.2\% at low skewness, both acceptable relative to consensus traffic.
Checkpoint traffic comprises 24.9\% of total at $\alpha=1.24$ and 69.8\% at $\alpha=0.99$, with the absolute checkpoint traffic remaining the same across different skewness factors. This reflects that higher skewness enables more operations per update budget, increasing consensus and retrieval traffic proportionally while checkpoint cost stays constant.
Overall, \sys achieves good network scalability with per-node traffic remaining nearly constant. %
\section{Related Works}
\label{sec:rel}

\paragraph{Reduce storage overhead in state machine replication.}
As previously mentioned in \cref{sec:bg}, horizontal scaling is a common practice in distributed storage and databases to handle large data volumes~\cite{gfs,bigtable,spanner}.
In these systems, data is partitioned into multiple shards which are individually replicated, similar to the sharded blockchains.

Several prior works have explored reducing storage overhead in state machine replication with crash-fault tolerance.
RS-Paxos~\cite{mu2014paxos} and CRaft~\cite{wang2020craft} utilize erasure coding to replicate the logs.
Only the leader node stores full logs, while the followers maintain coded fragments.
Racos~\cite{zarnstorff2024racos} further introduces leaderless replication to resolve the bottleneck on the leader, which achieve storage scalability.
The demand for Byzantine fault tolerance substantially complicates the challenge of reducing storage overhead.
In these systems, usually only the proposing node executes the transactions, while other nodes store the coded fragments of the logs to ensure the state is recoverable after the proposer fails.
Under Byzantine faults, the single execution result cannot be trusted, which brings the dilemma between security and storage scalability in \cref{sec:bg}.

\paragraph{Storage scalable solutions for cryptocurrencies.}
Vault~\cite{leung2018vault} proposes a sharded storage solution for account-based blockchains.
With the stateless approach, Utreexo~\cite{dryja2019utreexo} proposes a stateless blockchain for UTXO based cryptocurrencies, while \textsc{Edrax}~\cite{chepurnoy2018edrax} is capable for both UTXO and account-based models.
In this scenario, the clients hold the piece of state relevant to themselves (e.g., the UTXOs they own or their account balances) and the storage nodes in general stateless blockchains are unnecessary.
All these works are specific to cryptocurrencies.

\paragraph{Execute-order blockchains.}
Hyperledger Fabric~\cite{hyperledger} introduces the execute-order-validate paradigm for permissioned blockchains.
The execution is performed on few designated execution peers before ordering, and the updates are then ordered by the consensus layer and applied to the storage nodes.
While the consensus is secure, the execution must be performed by sufficient peers to ensure correctness, and the state must be stored by sufficient storage nodes to ensure availability.
Thus, Fabric also suffers from trading off storage scalability and security and can also benefit from \sys's state management approach.

\paragraph{Optimize state storage on nodes.}
Several methodologies aim to optimize state storage on nodes without altering the fundamental architecture of blockchains.
COLE~\cite{zhang2024cole} designs a column-based learned index for Blockchain systems and reports 94\% reduction in state storage overhead.
Vault~\cite{leung2018vault} also proposed an expiration mechanism to prune zero-balance accounts.
These techniques can be orthogonally applied to \sys to further reduce the storage overhead on nodes.

\paragraph{Scaling BFT protocols.}
Numerous research efforts have been dedicated to scaling BFT consensus protocols to accommodate a larger number of nodes.
SBFT~\cite{gueta2019sbft}, Narwhal~\cite{narwhal} and Autobahn~\cite{giridharan2024autobahn} improve scalability of transaction dissemination.
Some sharded BFT protocols like \textsc{Elastico}~\cite{luu2016secure}, \textsc{Ohie}~\cite{yu2020ohie} and Monoxide~\cite{wang2019monoxide} partition the consensus layer, so that each transaction is ordered by a subset of nodes.
Monoxide further partitions the execution with a cross shard relaying mechanism.
These works can improve the ordering (and execution for some works) scalability, but they do not reduce the storage overhead on each node.

\section{Conclusion}

In this work, we introduce \sys, a BFT protocol with scalable storage consumption without compromising security.
\sys's state management draws inspiration from distributed databases, while also inheriting the consistent ordering and execution mechanisms of blockchain protocols.
We posit that \sys broadens the potential adoption landscape for both blockchains and databases.
Building upon \sys, several avenues for future research can be pursued, such as developing collaborative databases across mutually distrusting organizations and designing execution models for on-chain big data processing.

\bibliography{paper}

\begin{thebibliography}{10}

\bibitem{shard-sybil}
Hafid Abdelatif, Senhaji~Hafid Abdelhakim, and Samih Mustapha.
\newblock A tractable probabilistic approach to analyze sybil attacks in
  sharding-based blockchain protocols, 2021.

\bibitem{amiri2021sharper}
Mohammad~Javad Amiri, Divyakant Agrawal, and Amr El~Abbadi.
\newblock Sharper: Sharding permissioned blockchains over network clusters.
\newblock In {\em Proceedings of the 2021 international conference on
  management of data}, pages 76--88, 2021.

\bibitem{hyperledger}
Elli Androulaki, Artem Barger, Vita Bortnikov, Christian Cachin, Konstantinos
  Christidis, Angelo De~Caro, David Enyeart, Christopher Ferris, Gennady
  Laventman, Yacov Manevich, et~al.
\newblock Hyperledger fabric: a distributed operating system for permissioned
  blockchains.
\newblock In {\em Proceedings of the thirteenth EuroSys conference}, pages
  1--15, 2018.

\bibitem{mysticeti}
Kushal Babel, Andrey Chursin, George Danezis, Anastasios Kichidis, Lefteris
  Kokoris{-}Kogias, Arun Koshy, Alberto Sonnino, and Mingwei Tian.
\newblock Mysticeti: Reaching the latency limits with uncertified dags.
\newblock In {\em 32nd Annual Network and Distributed System Security
  Symposium, {NDSS} 2025, San Diego, California, USA, February 24-28, 2025}.
  The Internet Society, 2025.

\bibitem{sui}
Sam Blackshear, Andrey Chursin, George Danezis, Anastasios Kichidis, Lefteris
  Kokoris-Kogias, Xun Li, Mark Logan, Ashok Menon, Todd Nowacki, Alberto
  Sonnino, Brandon Williams, and Lu~Zhang.
\newblock Sui lutris: A blockchain combining broadcast and consensus.
\newblock In {\em Proceedings of the 2024 on ACM SIGSAC Conference on Computer
  and Communications Security}, CCS '24, page 2606–2620, New York, NY, USA,
  2024. Association for Computing Machinery.

\bibitem{boneh2019batching}
Dan Boneh, Benedikt B{\"u}nz, and Ben Fisch.
\newblock Batching techniques for accumulators with applications to iops and
  stateless blockchains.
\newblock In {\em Annual International Cryptology Conference}, pages 561--586.
  Springer, 2019.

\bibitem{vbuterin2020rolluproadmap}
Vitalik Buterin.
\newblock A rollup-centric ethereum roadmap.
\newblock
  \url{https://ethereum-magicians.org/t/a-rollup-centric-ethereum-roadmap/4698},
  October 2020.
\newblock Fellowship of Ethereum Magicians (online forum post).

\bibitem{eip4844}
Vitalik Buterin, Dankrad Feist, Diederik Loerakker, George Kadianakis, Matt
  Garnett, Mofi Taiwo, and Ansgar Dietrichs.
\newblock Eip-4844: Shard blob transactions.
\newblock \url{https://eips.ethereum.org/EIPS/eip-4844}, February 2022.
\newblock Ethereum Improvement Proposal 4844, Online; accessed May 8, 2025.

\bibitem{pbft}
Miguel Castro and Barbara Liskov.
\newblock Practical byzantine fault tolerance.
\newblock In {\em Proceedings of the Third Symposium on Operating Systems
  Design and Implementation}, OSDI '99, page 173–186, USA, 1999. USENIX
  Association.

\bibitem{bigtable}
Fay Chang, Jeffrey Dean, Sanjay Ghemawat, Wilson~C Hsieh, Deborah~A Wallach,
  Mike Burrows, Tushar Chandra, Andrew Fikes, and Robert~E Gruber.
\newblock Bigtable: A distributed storage system for structured data.
\newblock {\em ACM Transactions on Computer Systems (TOCS)}, 26(2):1--26, 2008.

\bibitem{chepurnoy2018edrax}
Alexander Chepurnoy, Charalampos Papamanthou, Shravan Srinivasan, and Yupeng
  Zhang.
\newblock Edrax: A cryptocurrency with stateless transaction validation.
\newblock {\em Cryptology ePrint Archive}, 2018.

\bibitem{cooper2010benchmarking}
Brian~F Cooper, Adam Silberstein, Erwin Tam, Raghu Ramakrishnan, and Russell
  Sears.
\newblock Benchmarking cloud serving systems with ycsb.
\newblock In {\em Proceedings of the 1st ACM symposium on Cloud computing},
  pages 143--154, 2010.

\bibitem{spanner}
James~C Corbett, Jeffrey Dean, Michael Epstein, Andrew Fikes, Christopher
  Frost, Jeffrey~John Furman, Sanjay Ghemawat, Andrey Gubarev, Christopher
  Heiser, Peter Hochschild, et~al.
\newblock Spanner: Google’s globally distributed database.
\newblock {\em ACM Transactions on Computer Systems (TOCS)}, 31(3):1--22, 2013.

\bibitem{narwhal}
George Danezis, Lefteris Kokoris-Kogias, Alberto Sonnino, and Alexander
  Spiegelman.
\newblock Narwhal and tusk: a dag-based mempool and efficient bft consensus.
\newblock In {\em Proceedings of the Seventeenth European Conference on
  Computer Systems}, pages 34--50, 2022.

\bibitem{dang2019towards}
Hung Dang, Tien Tuan~Anh Dinh, Dumitrel Loghin, Ee-Chien Chang, Qian Lin, and
  Beng~Chin Ooi.
\newblock Towards scaling blockchain systems via sharding.
\newblock In {\em Proceedings of the 2019 international conference on
  management of data}, pages 123--140, 2019.

\bibitem{dryja2019utreexo}
Thaddeus Dryja.
\newblock Utreexo: A dynamic hash-based accumulator optimized for the bitcoin
  utxo set.
\newblock {\em Cryptology ePrint Archive}, 2019.

\bibitem{el2019blockchaindb}
Muhammad El-Hindi, Carsten Binnig, Arvind Arasu, Donald Kossmann, and Ravi
  Ramamurthy.
\newblock Blockchaindb: A shared database on blockchains.
\newblock {\em Proceedings of the VLDB Endowment}, 12(11):1597--1609, 2019.

\bibitem{eth-requirement}
Ethereum hardware requirements.
\newblock
  \url{https://geth.ethereum.org/docs/getting-started/hardware-requirements},
  2022.

\bibitem{eth-state}
How ethereum' state bloat problem is killing dapp performance.
\newblock
  \url{https://medium.com/\%40sohail_saifi/how-ethereums-state-bloat-problem-is-killing-dapp-performance-7f243a888459},
  2025.

\bibitem{ethereum_scaling_docs}
{Ethereum Foundation}.
\newblock Ethereum developer documentation: Scaling.
\newblock \url{https://ethereum.org/en/developers/docs/scaling/}, 2023.
\newblock Accessed: 2025-05-28.

\bibitem{notallstate2025}
{Ethereum Magicians community}.
\newblock Not all state is equal.
\newblock \url{https://ethereum-magicians.org/t/not-all-state-is-equal/25508/},
  2025.
\newblock Accessed: 2025-11-30.

\bibitem{feng2024slimarchive}
Hang Feng, Yufeng Hu, Yinghan Kou, Runhuai Li, Jianfeng Zhu, Lei Wu, and Yajin
  Zhou.
\newblock $\{$SlimArchive$\}$: A lightweight architecture for ethereum archive
  nodes.
\newblock In {\em 2024 USENIX Annual Technical Conference (USENIX ATC 24)},
  pages 1257--1272, 2024.

\bibitem{gfs}
Sanjay Ghemawat, Howard Gobioff, and Shun-Tak Leung.
\newblock The {Google} {File} {System}.
\newblock In {\em Proceedings of the {Nineteenth} {ACM} {Symposium} on
  {Operating} {Systems} {Principles}}, {SOSP} ’03, Bolton Landing, NY, USA,
  2003. Association for Computing Machinery.

\bibitem{algorand}
Yossi Gilad, Rotem Hemo, Silvio Micali, Georgios Vlachos, and Nickolai
  Zeldovich.
\newblock Algorand: Scaling byzantine agreements for cryptocurrencies.
\newblock In {\em Proceedings of the 26th symposium on operating systems
  principles}, pages 51--68, 2017.

\bibitem{giridharan2024autobahn}
Neil Giridharan, Florian Suri-Payer, Ittai Abraham, Lorenzo Alvisi, and Natacha
  Crooks.
\newblock Autobahn: Seamless high speed bft.
\newblock In {\em Proceedings of the ACM SIGOPS 30th Symposium on Operating
  Systems Principles}, pages 1--23, 2024.

\bibitem{gueta2019sbft}
Guy~Golan Gueta, Ittai Abraham, Shelly Grossman, Dahlia Malkhi, Benny Pinkas,
  Michael Reiter, Dragos-Adrian Seredinschi, Orr Tamir, and Alin Tomescu.
\newblock Sbft: A scalable and decentralized trust infrastructure.
\newblock In {\em 2019 49th Annual IEEE/IFIP international conference on
  dependable systems and networks (DSN)}, pages 568--580. IEEE, 2019.

\bibitem{shard-model}
Abdelatif Hafid, Abdelhakim~Senhaji Hafid, and Mustapha Samih.
\newblock New mathematical model to analyze security of sharding-based
  blockchain protocols.
\newblock {\em IEEE Access}, 7:185447--185457, 2019.

\bibitem{heimbach2024dissecting}
Lioba Heimbach, Quentin Kniep, Yann Vonlanthen, Roger Wattenhofer, and Patrick
  Z{\"u}st.
\newblock Dissecting the eip-2930 optional access lists.
\newblock In {\em International Conference on Financial Cryptography and Data
  Security}, pages 292--302. Springer, 2024.

\bibitem{hellings2021byshard}
Jelle Hellings and Mohammad Sadoghi.
\newblock Byshard: Sharding in a byzantine environment.
\newblock {\em Proceedings of the VLDB Endowment}, 14(11):2230--2243, 2021.

\bibitem{linearizability}
Maurice~P. Herlihy and Jeannette~M. Wing.
\newblock Linearizability: {A} {Correctness} {Condition} for {Concurrent}
  {Objects}.
\newblock {\em ACM Trans. Program. Lang. Syst.}, 12(3), July.

\bibitem{hong2024gridb}
Zicong Hong, Song Guo, Enyuan Zhou, Wuhui Chen, Huawei Huang, and Albert
  Zomaya.
\newblock Gridb: Scaling blockchain database via sharding and off-chain
  cross-shard mechanism.
\newblock {\em arXiv preprint arXiv:2407.03750}, 2024.

\bibitem{kannan1999simple}
Ravi Kannan, Prasad Tetali, and Santosh Vempala.
\newblock Simple markov-chain algorithms for generating bipartite graphs and
  tournaments.
\newblock {\em Random Structures \& Algorithms}, 14(4):293--308, 1999.

\bibitem{taocpv3}
Donald~E. Knuth.
\newblock {\em The art of computer programming, volume 3: (2nd ed.) sorting and
  searching}.
\newblock Addison Wesley Longman Publishing Co., Inc., USA, 1998.

\bibitem{kokoris2018omniledger}
Eleftherios Kokoris-Kogias, Philipp Jovanovic, Linus Gasser, Nicolas Gailly,
  Ewa Syta, and Bryan Ford.
\newblock Omniledger: A secure, scale-out, decentralized ledger via sharding.
\newblock In {\em 2018 IEEE symposium on security and privacy (SP)}, pages
  583--598. IEEE, 2018.

\bibitem{krol2021shard}
Micha{\l} Kr{\'o}l, Onur Ascigil, Sergi Rene, Alberto Sonnino, Mustafa
  Al-Bassam, and Etienne Rivi{\`e}re.
\newblock Shard scheduler: Object placement and migration in sharded
  account-based blockchains.
\newblock In {\em Proceedings of the 3rd ACM Conference on Advances in
  Financial Technologies}, pages 43--56, 2021.

\bibitem{lee2020replicated}
Jonathan Lee, Kirill Nikitin, and Srinath Setty.
\newblock Replicated state machines without replicated execution.
\newblock In {\em 2020 IEEE Symposium on Security and Privacy (SP)}, pages
  119--134. IEEE, 2020.

\bibitem{leung2018vault}
Derek Leung, Adam Suhl, Yossi Gilad, and Nickolai Zeldovich.
\newblock Vault: Fast bootstrapping for the algorand cryptocurrency.
\newblock {\em Cryptology ePrint Archive}, 2018.

\bibitem{leveldb}
{LevelDB} fast key-value storage library.
\newblock \url{https://github.com/google/leveldb}, 2011.

\bibitem{lin2025parallelevm}
Haoran Lin, Hang Feng, Yajin Zhou, and Lei Wu.
\newblock Parallelevm: Operation-level concurrent transaction execution for
  evm-compatible blockchains.
\newblock In {\em Proceedings of the Twentieth European Conference on Computer
  Systems}, pages 211--225, 2025.

\bibitem{luu2016secure}
Loi Luu, Viswesh Narayanan, Chaodong Zheng, Kunal Baweja, Seth Gilbert, and
  Prateek Saxena.
\newblock A secure sharding protocol for open blockchains.
\newblock In {\em Proceedings of the 2016 ACM SIGSAC conference on computer and
  communications security}, pages 17--30, 2016.

\bibitem{honeybadger}
Andrew Miller, Yu~Xia, Kyle Croman, Elaine Shi, and Dawn Song.
\newblock The honey badger of bft protocols.
\newblock In {\em Proceedings of the 2016 ACM SIGSAC conference on computer and
  communications security}, pages 31--42, 2016.

\bibitem{mu2014paxos}
Shuai Mu, Kang Chen, Yongwei Wu, and Weimin Zheng.
\newblock When paxos meets erasure code: Reduce network and storage cost in
  state machine replication.
\newblock In {\em Proceedings of the 23rd international symposium on
  High-performance parallel and distributed computing}, pages 61--72, 2014.

\bibitem{bitcoin}
Satoshi Nakamoto.
\newblock Bitcoin: A peer-to-peer electronic cash system.
\newblock \url{https://bitcoin.org/bitcoin.pdf}, 2008.
\newblock Accessed: 2025-05-20.

\bibitem{nodereal2022_bsc_cache_hit_rate}
{NodeReal}.
\newblock Bnb smart chain performance anatomy series: Chapter ii. 99\% cache
  hit rate.
\newblock
  \url{https://nodereal.io/blog/en/bnb-smart-chain-performance-anatomy-series-chapter-ii-99-cache-hit-rate/},
  2022.
\newblock Accessed: 2025-11-30.

\bibitem{MempoolUTXOReport2025}
orangesurf~(for Mempool~Research).
\newblock Utxo set report.
\newblock \url{https://research.mempool.space/utxo-set-report/}, 2025.
\newblock “The UTXO set contains 173 million UTXOs…”.

\bibitem{Reinheimer2020}
Paul Reinheimer.
\newblock A day in the life of the internet.
\newblock
  \url{https://wonderproxy.com/blog/a-day-in-the-life-of-the-internet/}, 10
  2020.
\newblock WonderProxy Blog.

\bibitem{rocksdb}
{RocksDB} persistent key-value store.
\newblock \url{https://rocksdb.org/}, 2012.

\bibitem{ruan2019fine}
Pingcheng Ruan, Gang Chen, Tien Tuan~Anh Dinh, Qian Lin, Beng~Chin Ooi, and
  Meihui Zhang.
\newblock Fine-grained, secure and efficient data provenance on blockchain
  systems.
\newblock {\em Proceedings of the VLDB Endowment}, 12(9):975--988, 2019.

\bibitem{ruan2021blockchains}
Pingcheng Ruan, Tien Tuan~Anh Dinh, Dumitrel Loghin, Meihui Zhang, Gang Chen,
  Qian Lin, and Beng~Chin Ooi.
\newblock Blockchains vs. distributed databases: Dichotomy and fusion.
\newblock In {\em Proceedings of the 2021 International Conference on
  Management of Data}, pages 1504--1517, 2021.

\bibitem{rsm}
Fred~B. Schneider.
\newblock Implementing {Fault}-{Tolerant} {Services} {Using} the {State}
  {Machine} {Approach}: {A} {Tutorial}.
\newblock {\em ACM Comput. Surv.}, 22(4), December 1990.

\bibitem{bullshark}
Alexander Spiegelman, Neil Giridharan, Alberto Sonnino, and Lefteris
  Kokoris-Kogias.
\newblock Bullshark: Dag bft protocols made practical.
\newblock In {\em Proceedings of the 2022 ACM SIGSAC Conference on Computer and
  Communications Security}, pages 2705--2718, 2022.

\bibitem{spiegelman2022bullshark}
Alexander Spiegelman, Neil Giridharan, Alberto Sonnino, and Lefteris
  Kokoris-Kogias.
\newblock Bullshark: The partially synchronous version.
\newblock {\em arXiv preprint arXiv:2209.05633}, 2022.

\bibitem{ssd-price}
Storage price trends.
\newblock \url{https://pcpartpicker.com/trends/price/internal-hard-drive/},
  2025.

\bibitem{state-bloat}
Solving state bloat.
\newblock
  \url{https://www.theblock.co/post/258960/solving-state-bloat?utm_source=chatgpt.com},
  2023.

\bibitem{tomescu2020aggregatable}
Alin Tomescu, Ittai Abraham, Vitalik Buterin, Justin Drake, Dankrad Feist, and
  Dmitry Khovratovich.
\newblock Aggregatable subvector commitments for stateless cryptocurrencies.
\newblock In {\em International Conference on Security and Cryptography for
  Networks}, pages 45--64. Springer, 2020.

\bibitem{wang2019monoxide}
Jiaping Wang and Hao Wang.
\newblock Monoxide: Scale out blockchains with asynchronous consensus zones.
\newblock In {\em 16th USENIX symposium on networked systems design and
  implementation (NSDI 19)}, pages 95--112, 2019.

\bibitem{wang2020craft}
Zizhong Wang, Tongliang Li, Haixia Wang, Airan Shao, Yunren Bai, Shangming Cai,
  Zihan Xu, and Dongsheng Wang.
\newblock $\{$CRaft$\}$: An erasure-coding-supported version of raft for
  reducing storage cost and network cost.
\newblock In {\em 18th USENIX Conference on File and Storage Technologies (FAST
  20)}, pages 297--308, 2020.

\bibitem{ethereum}
Gavin Wood et~al.
\newblock Ethereum: A secure decentralised generalised transaction ledger.
\newblock {\em Ethereum project yellow paper}, 151(2014):1--32, 2014.

\bibitem{xu2021slimchain}
Cheng Xu, Ce~Zhang, Jianliang Xu, and Jian Pei.
\newblock Slimchain: Scaling blockchain transactions through off-chain storage
  and parallel processing.
\newblock {\em Proceedings of the VLDB Endowment}, 14(11):2314--2326, 2021.

\bibitem{xu2022l2chain}
Zihuan Xu and Lei Chen.
\newblock L2chain: Towards high-performance, confidential and secure layer-2
  blockchain solution for decentralized applications.
\newblock {\em Proceedings of the VLDB Endowment}, 16(4):986--999, 2022.

\bibitem{hotstuff}
Maofan Yin, Dahlia Malkhi, Michael~K Reiter, Guy~Golan Gueta, and Ittai
  Abraham.
\newblock Hotstuff: Bft consensus with linearity and responsiveness.
\newblock In {\em Proceedings of the 2019 ACM symposium on principles of
  distributed computing}, pages 347--356, 2019.

\bibitem{yu2020ohie}
Haifeng Yu, Ivica Nikoli{\'c}, Ruomu Hou, and Prateek Saxena.
\newblock Ohie: Blockchain scaling made simple.
\newblock In {\em 2020 IEEE Symposium on Security and Privacy (SP)}, pages
  90--105. IEEE, 2020.

\bibitem{zamani2018rapidchain}
Mahdi Zamani, Mahnush Movahedi, and Mariana Raykova.
\newblock Rapidchain: Scaling blockchain via full sharding.
\newblock In {\em Proceedings of the 2018 ACM SIGSAC conference on computer and
  communications security}, pages 931--948, 2018.

\bibitem{zarnstorff2024racos}
Jonathan Zarnstorff, Lucas Lebow, Christopher Siems, Dillon Remuck, Colin Ruiz,
  and Lewis Tseng.
\newblock Racos: Improving erasure coding state machine replication using
  leaderless consensus.
\newblock In {\em Proceedings of the 2024 ACM Symposium on Cloud Computing},
  pages 600--617, 2024.

\bibitem{zhang2024cole}
Ce~Zhang, Cheng Xu, Haibo Hu, and Jianliang Xu.
\newblock $\{$COLE$\}$: A column-based learned storage for blockchain systems.
\newblock In {\em 22nd USENIX Conference on File and Storage Technologies (FAST
  24)}, pages 329--345, 2024.

\end{thebibliography}

\ifapx
\clearpage
\appendix
\section{Correctness Proof}

In this section, we prove that the state management layer can provide state availability and safety guarantees as stated in the paper.
The state availability guarantees mean that every $\fetch_{i}(k)$ operation at version $i$ will eventually return on all correct replica nodes, and state safety means it will return with the correct value of key $k$ at version $i$ on all these nodes except at most $f$ of them where it returns \forward.
The values of version $i$ are created with the $\bump_{i}(k_1 \mapsto v_1, k_2 \mapsto v_2, \ldots)$ operation, where the values of $k_1, k_2, \ldots$ are $v_1, v_2, \ldots$ respectively, and the values of other keys remain the same as version $i-1$.

Firstly, we examine the last checkpointed version $p$, which is the version that the second last checkpoint signal corresponds to.
We only need to consider safety of $\fetch_{i}(k)$ operations when $i \geq p$.
If $i < p$, the operation will return \forward according to the garbage collection rule, which is always safe. 
We show that this happens on at most $f$ correct replica nodes.
If there are more than $f$ correct replica nodes calling $\fetch_{i}(k)$ with $i < p$, they must be absent from the checkpointing procedure of version $p$, which means they did not send \votec messages in the round that checkpoints the version $p$.
However, the checkpoint signal for that round is formed with $2f+1$ matching \votec messages, which leads to a contradiction.
By ensuring that at most $f$ correct replica nodes call $\fetch_{i}(k)$ with $i < p$, the liveness of BFT system is not violated by returning \forward on these nodes, because there will be at least $f+1$ correct replica nodes \emph{not} returning \forward and thus able to proceed with execution, and the client can collect sufficient matching replies from them to commit.
In conclusion, the $\fetch_{i}(k)$ operations with $i < p$ ensure both availability and safety.

Next, we consider the $\fetch_{i}(k)$ operations where $i \geq p$.
According to the garbage collection rule, the replica nodes keep all the updates since version $p$ locally in the update table.
Thus, any $\fetch_{i}(k)$ operation where $k$ was updated after version $p$, i.e., appeared in some $\bump_{j}(\ldots)$ operation where $j > p$, will return correctly from the update table.
Otherwise, if $k$ was not updated after version $p$, $\fetch_{i}(k)$ should return the same value as $\fetch_{p}(k)$, so the value of $k$ in the checkpoint is the desired value, and the replica nodes send \retrieve requests to the storage nodes.
The state safety follows directly from the Merkle proof verification.
Now we prove the state availability of these \retrieve requests.

We first show that every reconfiguration attempt will eventually complete.
The last checkpoint has at least $f+1$ encoded chunks stored on distinct correct storage nodes.
This is because that the \pushss messages for the last checkpoint must have been acknowledged by at least $2f+1$ storage nodes, among which at least $f+1$ are correct nodes.
During a reconfiguration, the correct responsible storage nodes (after reconfiguration) send \getsc requests to all storage nodes.
They can detect faulty responses with the stored chunk hashes, and they are guaranteed to retrieve $f+1$ correct chunks from correct storage nodes and reconstruct the shard.
Thus, every reconfiguration attempt will eventually complete.
Because the placement is uniformly random, the probability that all $r$ responsible storage nodes are faulty in $x$ consecutive reconfigurations is at most $(\frac{f}{N})^{rx}$, which decreases exponentially with both $r$ and $x$.
After sufficiently many reconfigurations, with high probability at least one correct storage node will be assigned as responsible for the shard, and the \retrieve requests will eventually return with the correct value.
Thus, the $\fetch_{i}(k)$ operations with $i \geq p$ ensure both availability and safety. \fi
 
\end{document}